\documentclass[10pt,journal]{IEEEtran}

\usepackage{amsmath,amssymb}
\usepackage{theorem}
\usepackage{graphicx}
\usepackage{tabularx}
\usepackage{algorithm}
\usepackage{algorithmic}
\usepackage{bbm}
\usepackage{cite}
\usepackage{psfrag}
\usepackage{stfloats}
\usepackage[justification=centering]{caption}
\usepackage[caption=false,font=footnotesize]{subfig}
\newtheorem{theorem}{Theorem}
\newtheorem{lemma}{Lemma} 
\newtheorem{definition}{Definition}

\newtheorem{corollary}[theorem]{Corollary}
\makeatletter
\def\blfootnote{\gdef\@thefnmark{}\@footnotetext}
\makeatother

\begin{document}

\title{On the Capacity of Wireless Networks with Random Transmission Delay}
\author{Niv~Voskoboynik, Haim~H.~Permuter and~Asaf~Cohen}
\maketitle

\begin{abstract}
In this paper, we introduce novel coding schemes for wireless networks with random transmission delays. These coding schemes obviate the need for synchronicity, reduce the number of transmissions and achieve the optimal rate region in the corresponding wired model for both multiple unicast and multicast cases with up to three users under the equal rate constraint. The coding schemes are presented in two phases; first, coding schemes for line, star and line-star topologies with random transmission delays are provided. Second, any general topology with multiple bidirectional unicast and multicast sessions is shown to be decomposable into these canonical topologies to reduce the number of transmissions without rate redundancy. As a result, the coding schemes developed for the line, star and line-star topologies serve as building blocks for the construction of more general coding schemes for all networks. The proposed schemes are proved to be Real Time (RT) for wireless networks in the sense that they achieve the minimal decoding delay. With a negligible size header, these coding schemes are shown to be applicable
to unsynchronized networks, i.e., networks with random transmission delays. Finally, we demonstrate the applicability of these schemes by extensive simulations. The implementation of such coding schemes on a wireless network with random transmission delay can improve performance and power efficiency.
\end{abstract}

\blfootnote{This research was supported in part by the Israel Science Foundation (grant no. 684/11) and an ERC starting grant. This paper was presented in part at the 2014 IEEE International Symposium on Network Coding, Aalborg, Denmark. N. Voskoboynik and H. H. Permuter are with the Department of Electrical and Computer Engineering, Ben-Gurion University of the Negev, 84105, Beer-Sheva, Israel (email: voscoboy@post.bgu.ac.il; haimp@bgu.ac.il). A. Cohen is with the Department of Communication Systems Engineering, Ben-Gurion University of the Negev, 84105, Beer-Sheva, Israel (email: coasaf@bgu.ac.il).}

\begin{IEEEkeywords}
Multiple unicast, Network coding, Random delay, Wireless networks.
\end{IEEEkeywords}

\section{Introduction}\label{sec:introduction}
\IEEEPARstart{N}{etwork} Coding (NC) \cite{cite:flow,cite:field} is a networking technique used to better exploit the available bandwidth, use energy efficiently and increase the network's security. Unlike the traditional approach of routing, NC enables messages to be encoded and decoded at the intermediate nodes.

\par In wireless networks, NC allows the broadcast nature of the channel to be exploited and significant throughput gains to be achieved compared to traditional routing. Yet, in many practical scenarios, \emph{random transmission delays} may render existing NC solutions impractical. In this paper, we focus on energy efficient coding schemes for wireless networks that, on the one hand, exploit the benefits of NC in wireless networks and achieve the fundamental limits in several practical configurations but, on the other hand, \emph{obviate the need for synchronicity}, which is a key impediment in such networks.

\par
\begin{figure}[]
\centering
    \subfloat[Wireless model\label{fig:wireless}]{%
        \psfrag{a}[][][1]{$r$}
        \psfrag{b}[][][1]{$1$}
        \psfrag{c}[][][1]{$2$}
        \psfrag{d}[][][0.8]{$W_2^{(q)}$}
        \psfrag{e}[][][0.8]{$W_1^{(p)}{\oplus}W_2^{(q)}$}
        \psfrag{f}[][][0.8]{$W_1^{(p)}$}
        \psfrag{g}[][][0.8]{$\hat{W_2^{(0)}},\hat{W_2^{(1)}},...$}
        \psfrag{h}[][][0.8]{$W_1^{(0)},W_1^{(1)},...$}
        \psfrag{i}[][][0.8]{$W_2^{(0)},W_2^{(1)},...$}
        \psfrag{j}[][][0.8]{$\hat{W_1^{(0)}},\hat{W_1^{(1)}},...$}
        \includegraphics[width=2.5in]{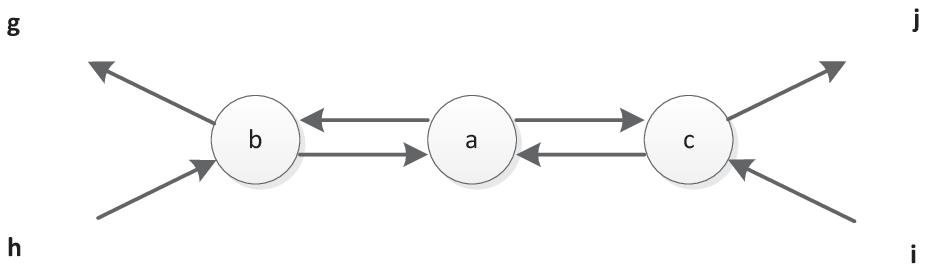}%
    }
    \hfill
    \subfloat[Wired model\label{fig:wired}]{%
        \psfrag{a}[][][1]{$r$}
        \psfrag{b}[][][1]{$1$}
        \psfrag{c}[][][1]{$2$}
        \psfrag{d}[][][0.8]{$W_2^{(q)}$}
        \psfrag{e}[][][0.8]{$W_1^{(p)}{\oplus}W_2^{(q)}$}
        \psfrag{f}[][][0.8]{$W_1^{(p)}$}
        \psfrag{g}[][][0.8]{$\hat{W_2^{(0)}},\hat{W_2^{(1)}},...$}
        \psfrag{h}[][][0.8]{$W_1^{(0)},W_1^{(1)},...$}
        \psfrag{i}[][][0.8]{$W_2^{(0)},W_2^{(1)},...$}
        \psfrag{j}[][][0.8]{$\hat{W_1^{(0)}},\hat{W_1^{(1)}},...$}
        \psfrag{p}[][][0.8]{$r'$}
        \includegraphics[width=2.5in]{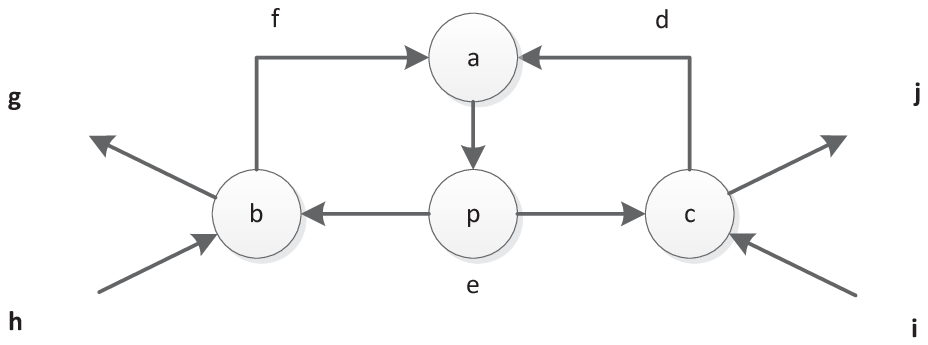}%
    }
    \caption{Schematic illustrations of \protect\subref{fig:wireless} wireless and \protect\subref{fig:wired} wired setups, where nodes $1$ and $2$ exchange their information through node $r$.}
    \label{fig:wired_to_wireless}
\end{figure}

 The following example demonstrates the difficulty of implementing a NC scheme for a network with random transmission delay. Consider two users, $1$ and $2$, that exchange packets $\{W_1^{(0)},...,W_1^{(t)}\}$ and $\{W_2^{(0)},...,W_2^{(t)}\}$ through a relay, $r$, in a packet-based communication scheme (Fig. \ref{fig:wireless}). The links can carry one packet per time unit, and each transmission has a discrete random delay. We assume that at time instant $t'$ the relay, $r$, receives a packet-pair $W_1^{(p)},W_2^{(q)}$, where $p$ and $q$ are arbitrary integers that are smaller than $t$. Using simple routing, the relay transmits each packet separately, and therefore, the minimum number of transmissions required by node $r$ is two for each packet-pair $W_1^{(p)},W_2^{(q)}$. In contrast, the NC approach suggests that node $r$ could encode two packets and broadcast the encoded packet by exploiting the wireless medium, e.g., $W_1^{(p)}\oplus{W_2^{(q)}}$, where $\oplus$ is a bitwise XOR operation. However, to decode the encoded packet at both nodes $1$ and $2$, the values of $p$ and $q$ are required. A straightforward NC implementation, therefore, cannot be decoded successfully. A feasible solution, which is to code only within a single \emph{generation} of packets and decode each generation separately, may incur an unacceptable delay. Of course, the problem becomes even more involved with the inclusion of multiple sources and terminals, a scenario that introduces several independent, bi-directional sessions. A key goal in our work, therefore, is to find coding schemes that not only exploit NC to minimize the number of transmissions, but that also have a feasible decoding process with \emph{minimum decoding delay} and a small overhead.

\begin{figure*}[]
\centering
    \subfloat[Multicast model\label{fig:model_multicast}]{%
        \psfrag{a}[][][0.8]{$1$}
        \psfrag{b}[][][0.8]{$2$}
        \psfrag{c}[][][0.8]{$3$}
        \psfrag{u}[][][0.8]{$\hat{W_2^{(t)}}$}
        \psfrag{v}[][][0.8]{$\hat{W_3^{(t)}}$}
        \psfrag{w}[][][0.8]{${W_1^{(t)}}$}
        \psfrag{q}[][][0.8]{$\hat{W_1^{(t)}}$}
        \psfrag{p}[][][0.8]{${W_2^{(t)}}$}
        \psfrag{r}[][][0.8]{${W_3^{(t)}}$}
        \includegraphics[width=2.5in]{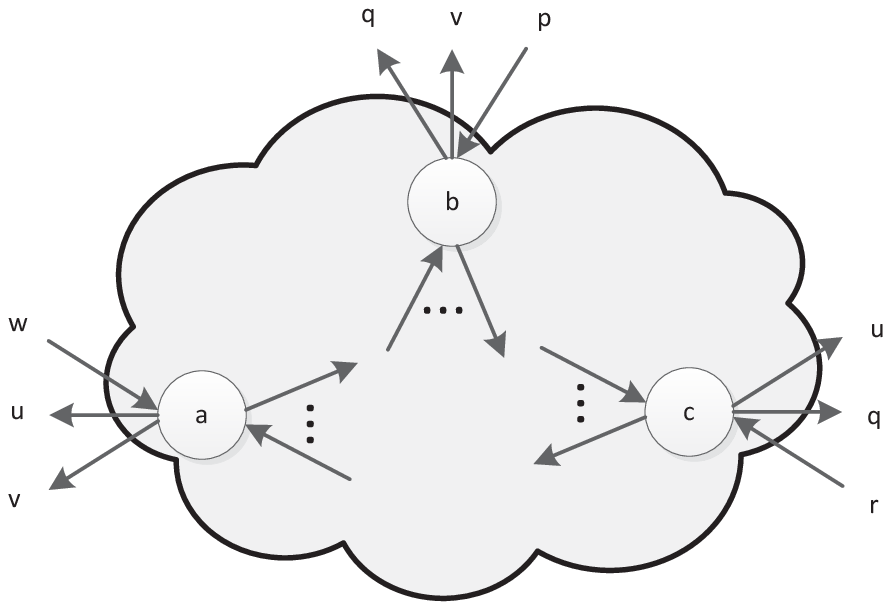}%
    }
    \quad\quad\quad\quad
    \subfloat[Multiple unicast model\label{fig:model_unicast}]{%
        \psfrag{a}[][][0.8]{$1$}
        \psfrag{b}[][][0.8]{$2$}
        \psfrag{c}[][][0.8]{$3$}
        \psfrag{w}[][][0.8]{${W_{1\rightarrow{2}}^{(t)}}$}
        \psfrag{v}[][][0.8]{${W_{1\rightarrow{3}}^{(t)}}$}
        \psfrag{p}[][][0.8]{${W_{2\rightarrow{1}}^{(t)}}$}
        \psfrag{q}[][][0.8]{${W_{2\rightarrow{3}}^{(t)}}$}
        \psfrag{u}[][][0.8]{${W_{3\rightarrow{1}}^{(t)}}$}
        \psfrag{r}[][][0.8]{${W_{3\rightarrow{2}}^{(t)}}$}
        \includegraphics[width=2.5in]{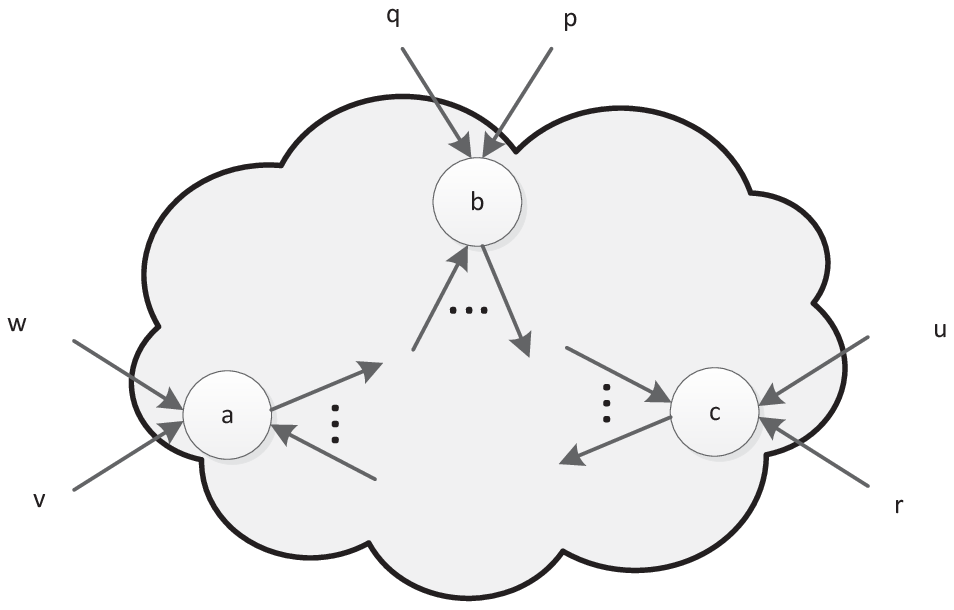}%
    }
    \caption{Schematic illustration of the multicast model \protect\subref{fig:model_multicast} and the multiple unicast model \protect\subref{fig:model_unicast}, where $W_i^{(t)}$ is a message that node $i$ generates at time instant $t$ and that is destined for all the other source nodes, and $W_{i\rightarrow{j}}^{(t)}$ is a message that node $i$ generates at time instant $t$ and that is destined only for node $j$.}
    \label{fig:model}
\end{figure*}

\par We first define the wireless network model to be used throughout this work. Following \cite{cite:wired_model}, we transform a given wireless network into a corresponding wired model (Fig. \ref{fig:wired_to_wireless}). Namely, we define a \emph{bidirectional graph} with random transmission delays, translating the criterion for optimality from minimum transmissions (in a wireless model) into maximum rates (in a wired model). For example, in the wired model of Fig. \ref{fig:wired}, using simple routing yields a rate of $R_1+R_2\leq{C}$, where $R_i$ is the rate of source $i$ and $C$ is the capacity of each link, while using NC yields the maximum transmission rate of $R_i\leq{C}$, ${i}\in\{1,2\}$. Our objective is thus to derive coding schemes for the wireless model that \emph{achieve the capacity rate region in the corresponding wired model} with minimum delay.

\par A key network characteristic that significantly impacts our ability to give tight results and optimal NC schemes is the \emph{demand structure}. For example, in a \emph{multicast} scenario \cite{cite:code_multicast,cite:random}, all terminal nodes wish to decode all sources. For this case, practical and rate-optimal solutions exist under several network models. However, in a \emph{multiple unicast} scenario, where independent source-destination pairs wish to communicate, the problem is much less tractable, and the \emph{general case is still unsolved}. In fact, the authors of \cite{cite:equivalent} showed that any acyclic directed network (with general demand structure) has an equivalent multiple unicast network, conferring significant interest on the study of such networks. Specifically, unlike the multicast case, linear NC \cite{cite:linear} fails to achieve the capacity region in this setting \cite{cite:insufficiency}. A few special cases are the capacity region when \emph{only XOR} operations are permitted, given in \cite{cite:unicast_lo}, and a coding scheme for \emph{three unicast sessions} that achieves a rate of half the minimum cut using the interference alignment approach, which was presented in \cite{cite:interference}. Our setup, on the other hand, presents three \emph{bidirectional} sessions. The case of three unicast sessions was also studied in \cite{cite:3source}, where lower bounds on the connectivity of the network, which allows a unity rate, were introduced. Here, however, we give a coding scheme that achieves capacity without assuming anything about the connectivity of the network.

\par Thus, we continue to explore the case of three unicast sessions, but in our case, the three sessions are bidirectional and experience random delays. Additionally, similar to the opportunistic coding approach that was presented in \cite{cite:xor,cite:coding_aware_routing}, we decrease the number of transmissions. Motivated by the conjecture in \cite{cite:conjecture}, which claimed that in undirected graphs there is no advantage to using NC, we show that despite the fact that the use of NC does not result in rate improvement, we introduce other advantages, such as a reduction in the number of transmissions in the corresponding bidirectional wired model. Specifically, in that model, which is a directed graph, we achieve the capacity rate region where simple routing schemes failed. We can now summarize our main contributions.

\subsection*{Main Contributions:}
We study in detail the two demand structures above under the suggested wireless model with random transmission delays. Specifically, we consider bidirectional multicast and multiple unicast \cite{cite:unicast1,cite:unicast2,cite:NCvsROUTING}, both for up to three users. In the first model (Fig. \ref{fig:model_multicast}), three users $1,2$ and $3$ exchange messages in a \emph{bidirectional multicast} manner through a wireless network. Multicast transmission is a widely used networking technique in which a message is sent to a set of receivers, an example of which is a video conference between three users. In the second model, each user generates \emph{two different messages}, one for each of the remaining users, i.e., it exchanges independent messages with two different users in a bidirectional manner through a wireless network (Fig. \ref{fig:model_unicast}). For an example of the multiple unicast case, consider a messaging application with several users, where each user communicates with its partners, but each message is addressed to a specific partner. The two paradigms, multicast and multiple unicast, are used as sub-networks in many different communication networks, such as wireless ad-hoc networks, server-client communication in cloud computing and networks of optical fibers.

\par For both demand structures, we \emph{achieve the capacity rate regions} in corresponding wired models, where each two-way communication is carried out at the same rate. Additionally, our coding schemes are shown to be Real Time (RT) NC, which we define as a NC scheme that allows decoding with minimum delay. Such a scheme is suitable for several applications, such as video conferencing and instant messaging.

\par In the multicast scenario, the benefits of the coding schemes are in their practicality and applicability to networks with random delays. In the multiple unicast scenario, which is generally open, we also \emph{extend the current state of the art as to when network coding is optimal and what are the achievable rates}.

\par Our coding schemes are based on a modular approach. We begin by providing simple constructions for line and star topologies (e.g., \cite{cite:line_dep,cite:node_constrained}), after which we use graph-theoretic tools to show how to decompose general networks into the above building blocks. Specifically, constructive algorithms are given to show that wireless networks with random transmission delays can be decomposed into line and star topologies as building blocks \emph{without rate redundancy} and RT coding schemes can be deployed with a \emph{small overhead and minimum transmissions}.

\par The rest of the paper is organized as follows. In Section \ref{sec:setup}, we present the network model. In Section \ref{sec:preliminaries}, we outline the preliminaries of the coding schemes for the line, star and line-star topologies and then use these coding schemes in Sections \ref{sec:multicast} and \ref{sec:unicast} as building blocks to derive a coding scheme for a general wireless network with multicast and multiple unicast sessions, respectively. Simulation results are given in Section \ref{sec:simulation} and then some extensions are presented in Section \ref{sec:extensions}. Finally, in Section \ref{sec:conclusion} we summarize the paper with our conclusions.

\section{Notation and Problem Setup}\label{sec:setup}
A \emph{bidirectional wireless network with delays} is defined as a directed graph, $\mathcal{G(V,E)}$, where $\mathcal{V}=\{1,...,M\}$ is a set of nodes and $\mathcal{E}\subseteq{[1,...,M]\times[1,...,M]}$ is a set of bidirectional edges. Each edge $(i,j)\in\mathcal{E}$ represents a directed link from node $i$ to node $j$ with a capacity of $C$ bits per time unit. Since the edges are bidirectional, each edge $(i,j)\in\mathcal{E}$ induces a corresponding edge $(j,i)\in\mathcal{E}$ with the same capacity. Additionally, we consider a set of source nodes $\mathcal{S}\subseteq\mathcal{V}$.

\par
\begin{figure}[h]
\centering
    \subfloat[Wireless\label{fig:wireless2}]{
    \psfrag{b}[][][1]{$i$}
    \psfrag{a}[][][1]{$j$}
    \includegraphics[scale=0.8]{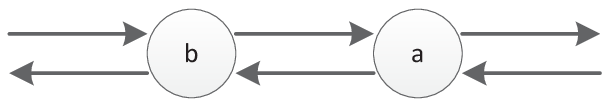}%
    }
    \hfil
    \subfloat[Wired\label{fig:wired2}]{
    \psfrag{a}[][][1]{$3$}
    \psfrag{b}[][][1]{$i$}
    \psfrag{a}[][][1]{$j$}
    \psfrag{d}[][][1]{$j'$}
    \psfrag{c}[][][1]{$i'$}
    \includegraphics[scale=0.8]{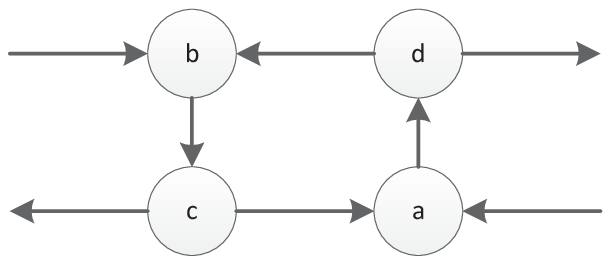}%
    }
    \caption{Diagram showing the conversion of a graph $\mathcal{G(V,E)}$ in \protect\subref{fig:wireless2} into a graph $\mathcal{G'(V',E')}$ in
            \protect\subref{fig:wired2}, $i,j\in\mathcal{V\setminus{S}}$.}
    \label{fig:node_constrained}
\end{figure}

 Next, we present a model that allows us to explore the broadcast ability of the wireless medium. Therefore, we introduce an equivalent directed graph, $\mathcal{G'(V',E')}$, with the same set of source nodes $\mathcal{S}\subseteq\mathcal{V'}$, by splitting each relay node $i\in\mathcal{V\setminus{S}}$ into two nodes $\{i,i'\}\subset\mathcal{V'}$ (Fig. \ref{fig:node_constrained}). Each pair of directed edges $(i,j)$ and $(j,i)$ in $\mathcal{E}$ corresponds to a pair of new directed edges, one entering $i$, $(j',i)\in\mathcal{E'}$, with capacity $C$ and another leaving $i'$, $(i',j)\in\mathcal{E'}$, with the same capacity. In addition, there is an edge directed from $i$ to $i'$ with capacity $C$ that models the broadcast constraint at relay node $i$.

\par In addition, we introduce the following notations:
\begin{itemize}
\item $W_i^{(t)}$ is the multicast message of source $i$ at time $t$. We assume each message is distributed uniformly over $\{1,...,2^{R_i}\}$.
\item $W_{i\rightarrow{j}}^{(t)}$ is the unicast message of source $i$ intended for node $j$ at time $t$. Again, we assume each message is distributed uniformly over $\{1,...,2^{R_{i\rightarrow{j}}}\}$.
\item $\mathcal{W}_i^t=\{W_{i}^{(0)},W_{i}^{(1)},...,W_{i}^{(t)}\}$ represents the set of messages that was produced by source $i$ up to time $t$. Similarly, $\mathcal{W}_{i\rightarrow{j}}^t=\{W_{i\rightarrow{j}}^{(0)},W_{i\rightarrow{j}}^{(1)},...,W_{i\rightarrow{j}}^{(t)}\}$.
\item $X_i^{(t)}$ represents the binary vector transmitted {\emph{on all the edges}} leaving node $i$ $\{(i,j) : j=1,...,M, (i,j)\in\mathcal{E'}\}$ at discrete time $t$.
\end{itemize}

\par We assume messages at negative times equal zero, i.e., $W_{i\rightarrow{j}}^{(t)}=0$ and $W_i^{(t)}=0$, $\forall{t}<0$ and $\forall{i,j}\in\mathcal{S}$. Our model consists of random transmission delays, i.e., $X_{i}^{(t)}$ is sent from node $i$ to node $j$ in time slot $t$ and yet received by node $j$ after an arbitrary discrete delay, $d^{(t)}_{i,j}$, $(i,j)\in\mathcal{E'}$. The delay is assumed to be bounded by $D$, i.e., $d^{(t)}_{i,j}\leq{D}$, $\forall{(i,j)\in\mathcal{E'}}$ and $\forall{t}$. We assume there is no delay in the node processors, i.e., the transmission over the edge $(i,i')\in\mathcal{E'}$ has no delay, for all $i\in\mathcal{V'}\setminus\mathcal{S}$. The outgoing transmission from every node at any particular time instant $t$ is a function of the incoming transmissions to that node at earlier time instants and of its own messages. Throughout the paper, we use the operators floor $\lfloor{\cdot}\rfloor$ and ceiling $\lceil{\cdot}\rceil$.

\par We denote by $\mathcal{C}_{i;j}$ the value of the minimal cut between nodes $i$ and $j$ in $\mathcal{G'}$. Symmetry between source nodes $i$ and $j$ induces that $\mathcal{C}_{i;j}=\mathcal{C}_{j;i}$. Similarly, we denote by $\mathcal{P}_{i;j}$ a subset of the set of disjoint paths from node $i$ to node $j$ in $\mathcal{G'}$, where it follows that $|\mathcal{P}_{i;j}|\leq\frac{\mathcal{C}_{i;j}}{C}$. In addition, we define a maximum distance in a graph $\mathcal{G'(V',E')}$ as $L=\max{\mathcal{P}_{i;j}}$, where the maximum is taken with respect to all $\mathcal{P}_{i;j}$, ${i,j}\in\mathcal{S}$. The maximum distance excludes the edges $(i,i')\in\mathcal{E'}$, $\forall{i}\in\mathcal{V'}\setminus\mathcal{S}$ since they have no delay. Furthermore, the diameter of the graph is the maximum distance in a network with $\mathcal{S}=\mathcal{V}$. Additionally, we will use the following definitions throughout the paper.

\begin{definition}[Achievable Rate]
A coding scheme of rates $(R_{i\rightarrow{j}},R_i,R_j)$ is said to be achievable if every node $j\in\mathcal{S}$ receives messages that are destined for it, i.e., $W_{i\rightarrow{j}}^{(t)}$ and $W_{i}^{(t)}$, $i\in\mathcal{S}\setminus\{j\}$, without error and with delay of at most $LD+c$ for all $t$, where $c$ is some constant that is independent of $D$.
\end{definition}

\begin{definition}[Equal Rate Capacity Region]
The capacity region under the equal rate constraint is defined as the closure of the set of all achievable rate tuples $(R_{i\rightarrow{j}},R_{j\rightarrow{i}},R_i,R_j)$, $i,j\in\mathcal{S}$, with the demands $R_{i\rightarrow{j}}=R_{j\rightarrow{i}}$ and $R_i=R_j$.
\end{definition}

\begin{definition}[Real Time]
A coding scheme is said to be RT for a graph $\mathcal{G'(V',E')}$ with a maximum distance $L$ and an arbitrary delay bounded by $D$ if every node $j\in\mathcal{S}$ decodes all messages up to time $t$, $\mathcal{W}_i^{t},\mathcal{W}_{i\rightarrow{j}}^{t}$, with a maximum delay of $LD+t+c$, $\forall{t}$, where $c$ is some constant that is independent of $D$.
\end{definition}

The minimum delay of any coding scheme on a path with length $L$ is bounded by $LD$, where a minimum delay is the upper bound of the worst case decoding delay of a coding scheme. An RT coding scheme decodes a new message in each time slot after an initialization duration of $LD+c$, i.e., a RT coding scheme achieves the minimum decoding delay. For comparison, a scheme based on random linear network coding (RLNC), e.g., \cite{cite:practical}, is not RT since it has to accumulate multiple transmissions before decoding is possible. Additionally, in a RT coding scheme the decoding delay is independent of the minimum cut between the source and the sink, where in RLNC this parameter has a huge impact on the decoding delay.

\par The goal is to find a RT coding scheme for a graph $\mathcal{G'(V',E')}$ that achieves the capacity region under an equal rate assumption. Furthermore, the coding scheme presented here requires only an overhead (as a header) of the order of $O(\log_2{h})$ bits to sperate the network into the building blocks, while a RLNC scheme requires $O(h)$ bits to transmit the global encoding vector, where $h$ is the minimal cut of separating one source from the network.

\section{Preliminaries}\label{sec:preliminaries}
In this section, we describe the key concepts of the coding scheme for the line, star and line-star topologies. We later use these schemes as building blocks for more complex networks.

\subsection{Line Topology}\label{sec:line}
\begin{figure}[!h]
\centering
    \psfrag{a}[][][1]{$1$}
    \psfrag{d}[][][1]{$2$}
    \psfrag{c}[][][1]{$2'$}
    \psfrag{e}[][][1]{$3$}
    \psfrag{f}[][][1]{$3'$}
    \psfrag{g}[][][1]{$M$}
    \includegraphics[width=2.5in]{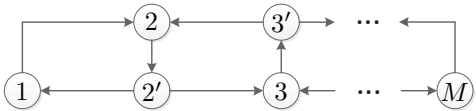}
    \caption{Graph $\mathcal{G'(V',E')}$ of a line topology, where $\mathcal{S}=\{1,M\}$.}
    \label{fig:line}
\end{figure}

A line topology of $M$ nodes is defined as a network $\mathcal{G'(V',E')}$ of two source nodes $\mathcal{S}=\{1,M\}$ that exchange messages $\mathcal{W}_1^{t}$ and $\mathcal{W}_M^{t}$ through a line of nodes (Fig. \ref{fig:line}). The coding scheme for this topology was first derived in \cite{cite:line}. For completeness and since we use it extensively later, we now present a sketch of the scheme. The main result for a line topology is summarized in the following theorem.
\begin{theorem}\label{theorem:line}
For any line topology with an arbitrary delay bounded by $D$, there exists a RT coding scheme that achieves the equal rate capacity, which is $C$. Furthermore, the coding scheme has a decoding delay of at most $LD$ and it includes a fixed header per transmission of $2\lceil\log_{2}2D\rceil$ bits, independent of $C$.
\end{theorem}

Since the minimal cut $\mathcal{C}_{1;M}=C$, we obtain that the equal rate upper bound is $R\leq{C}$. Next, we prove Theorem \ref{theorem:line} for the case where the delay per each transmission $d^{(t)}_{i,j}$ at time instant $t$ is fixed and equal to one, $\forall{(i,j)}\in\mathcal{E'}$ and later we provide a proof for the case where $d^{(t)}_{i,j}$ is an arbitrary integer, yet it is bounded by $D$. Note that for a case of unbounded delay, a retransmission policy is required, i.e., messages that were lost have to be retransmitted, however, in our case no retransmission is necessary.

\begin{IEEEproof}
{\it{Coding scheme:}} The transmission from source $i\in\mathcal{S}$ is a linear combination in $\mathbb{F}_{2^C}$, as follows
\begin{equation}\label{eq:line1}
X_i^{(t)}=k_1W_1^{(t-(i-1))}+k_MW_M^{(t-(M-i))},
\end{equation}
where $k_j\neq{0}$ is a fixed coefficient that is known a priori over a field $\mathbb{F}_{2^C}$, $j\in\{1,M\}$. Every relay node $r\in\mathcal{V'}\setminus\mathcal{S}$ generates
\begin{equation}\label{eq:line2}
X_r^{(t)}=X_{r+1}^{(t-1)}+X_{r-1}^{(t-1)}+X_r^{(t-2)}.
\end{equation}
We now show that the coding scheme defined in (\ref{eq:line1}) and (\ref{eq:line2}) implies that
\begin{equation}
X_r^{(t)}=k_1W_1^{(t-(r-1))}+k_MW_M^{(t-(M-r))}\label{eq:line1_gen},
\end{equation}
by induction over $t$. At time $t=0$, this claim is trivial, and therefore, we assume that up to time $t$ the claim holds. We prove the claim for $t+1$, where
\begin{align}
X_r^{(t+1)}&=X_{r+1}^{(t)}+X_{r-1}^{(t)}+X_r^{(t-1)}\nonumber\\
&=k_1W_1^{(t-r)}+k_MW_M^{(t+1-(M-r))}+k_1W_1^{(t-(r-2))}\nonumber\\
&\quad +k_MW_M^{(t-1-(M-r))}+k_1W_1^{(t-r)}+k_MW_M^{(t-1-(M-r))}\nonumber\\
&=k_1W_1^{(t+1-(r-1))}+k_MW_M^{(t+1-(M-r))},\nonumber
\end{align}
thus establishing the claim. Note that a field $\mathbb{F}_2^C$ is also applicable, but since we use it later as a building block, we chose $\mathbb{F}_{2^C}$.

\par {\it{Decoding:}} For each time instant $t$, each decoder $i\in\mathcal{S}$ subtracts the message $k_iW_i^{(t-1)}$ to decode the information $W_j^{(t-(M-1))}$, $j\in\mathcal{S}\setminus\{i\}$.

\par We verify by induction over $t$ that this NC scheme enables node $1$ to decode the message $W_M^{(t)}$ and node $M$ to decode the message $W_1^{(t)}$ with delay $M-1$. At time $t=0$, the claim is true, since messages at negative times are empty. By the inductive assumption, up to time $t$, node $1$ can recover $\mathcal{W}_M^{t-(M-1)},\mathcal{W}_1^{t}$ and node $M$ can recover $\mathcal{W}_1^{t-(M-1)},\mathcal{W}_M^{t}$. At time $t+1$, node $1$ obtains the transmission $k_1W_1^{(t-1)}+k_MW_M^{(t-(M-2))}$ and hence, it is able to decode message $W_M^{(t-(M-2))}$. Similarly, at time $t+1$, node $M$ obtains the transmission $k_1W_1^{(t-(M-2))}+k_MW_M^{(t-1)}$ and is able to decode $W_1^{(t-(M-2))}$, thereby establishing the claim. Furthermore, we obtain that the maximum delay for decoding messages $\mathcal{W}^t_i$ at source node $j$ is $L+t$ for all $t$, where $L=M-1$ and $i,j\in\mathcal{S}$. Therefore, this LNC scheme is, indeed, a RT coding scheme.
\end{IEEEproof}
Now we present the proof of Theorem \ref{theorem:line} for the case where $d^{(t)}_{i,j}$ is an arbitrary integer that can be different for each transmission over the edge $(i,j)\in\mathcal{E'}$ but that is bounded by $D$.

\begin{figure}[!h]
\centering
    \psfrag{a}[][][1]{$p$}
    \psfrag{b}[][][1]{$q$}
    \psfrag{c}[][][1]{$Data$}
    \includegraphics[width=2.5in,height=0.2in]{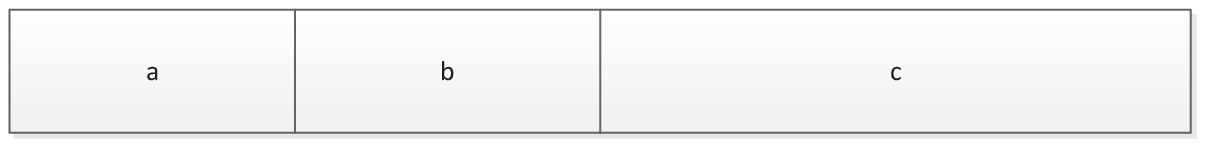}
    \caption{Packet header of an unsynchronized coding scheme for a line topology.}
    \label{fig:packet_line}
\end{figure}

\begin{IEEEproof}
We first note that Equation (\ref{eq:line2}) is no longer applicable for the unsynchronized case, and therefore, we introduce a new coding scheme for this case.
\par{\it{Coding scheme:}} The transmission from each node $r\in\mathcal{V'}$ is a linear combination in $\mathbb{F}_{2^C}$, as follows
\begin{equation}\label{eq:line_random}
X_{r}^{(t)}=k_1W_1^{(p)}+k_MW_M^{(q)},
\end{equation}
where $k_i\neq{0}$ is a coefficient over the field $\mathbb{F}_{2^C}$, and $W_1^{(p)}$ and $W_M^{(q)}$ are some arbitrary messages from the sets $\mathcal{W}_1^t$ and $\mathcal{W}_M^t$, respectively.

\par{\it{Decoding:}} For each incoming transmission, each decoder $i\in\mathcal{S}$ subtracts the message $k_iW_i^{(p)}$ to decode the information $W_j^{(q)}$, $j\in\mathcal{S}\setminus\{i\}$, for some arbitrary integers $p$ and $q$.

\par We verify that each node is able to produce a transmission in the form of (\ref{eq:line_random}) at any particular time instant $t$. First, we add two indices, each of length $\lceil{\log_2{2D}}\rceil$, as metadata to each transmission (Fig. \ref{fig:packet_line}). This header represents the messages from the sets $\mathcal{W}_1^t$ and $\mathcal{W}_M^t$ that were encoded. Second, we  demand that each node will decode and store the messages. We argue that the set of messages $\mathcal{W}_1^{p}$ that node $r$ holds at time instant $t$ is a subset of the set of messages $\mathcal{W}_1^{p'}$ held by node $r-1$. Similarly, node $r$ holds $\mathcal{W}_M^{q}$, which is a subset of the messages $\mathcal{W}_M^{q''}$ that node $r+1$ holds. We verify this claim using induction. At time $t=0$, the claim is trivial, and thus, we inductively assume that up to time $t$ the claim is true. We demonstrate the claim subsequently, where node $r$ obtains a transmission $X_{r-1}^{(t')}$, e.g., $k_1W_1^{(p')}+k_MW_M^{(q')}$ from node $r-1$. Since by the inductive assumption node $r$ has message $W_M^{(q')}$, it is able to decode message $W_1^{(p')}$; likewise, from the transmission from node $r+1$, e.g., $X_{r+1}^{(t'')}=k_1W_1^{(p'')}+k_MW_M^{(q'')}$, node $r$ is able to decode message $W_M^{(q'')}$, and hence, the claim is established. We summarize with the inference that node $r\in\mathcal{V'}$ is able to produce transmissions in the form of (\ref{eq:line_random}) at any particular time instant $t$ from messages it holds, independent of $d^{(t)}_{i,j}$, $\forall(i,j)\in\mathcal{E'}$. The maximum delay in this case is also bounded by $DL+t$ for all $t$, where $L=M-1$, and therefore, this coding scheme is a RT coding scheme.
\end{IEEEproof}

\par In the corresponding wireless model, i.e., a line network $\mathcal{G(V,E)}$, applying this coding scheme requires $|\mathcal{V}|-2$ transmissions by the relay nodes for each pair of messages $W_1^{(t)},W_M^{(t)}$, where using a simple routing scheme yields $2(|\mathcal{V}|-2)$ transmissions. Hence, by exploiting the broadcast ability of the wireless medium, we obtain an energy efficient coding scheme for the wireless setting.

\subsection{Star Topology}\label{sec:star}
\begin{figure}[!h]
\centering
    \psfrag{a}[][][1]{$1$}
    \psfrag{b}[][][1]{$2$}
    \psfrag{c}[][][1]{$3$}
    \psfrag{d}[][][1]{$4$}
    \psfrag{e}[][][1]{$4'$}
    \psfrag{f}[][][1]{$5$}
    \psfrag{g}[][][1]{$5'$}
    \includegraphics[width=2.5in,height=1in,keepaspectratio]{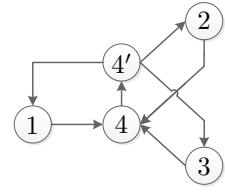}
    \caption{Graph $\mathcal{G'(V',E')}$ of a star topology, where $\mathcal{S}=\{1,2,3\}$.}
    \label{fig:star}
\end{figure}

A star topology of three source nodes is defined by a network $\mathcal{G'(V',E')}$, where $\mathcal{S}=\{1,2,3\}$ and all the source nodes in $\mathcal{S}$ try to communicate through a single relay node, $4$ (Fig. \ref{fig:star}). Our main result for the star topology is summarized in the following theorem.

\begin{theorem}\label{theorem:star}
For a star topology with an arbitrary delay bounded by $D$, there exists a RT coding scheme that achieves the equal rate capacity, which is $\frac{C}{2}$. Furthermore, the coding scheme includes a fixed header per transmission of $3\lceil\log_22D\rceil+1$ bits.
\end{theorem}

Since the minimal cut $\mathcal{C}_{i,j;l}=C$, we obtain that $R_i+R_j\leq{C}$, $\forall{i,j,l}\in\mathcal{S}$ and the upper bound of the equal rate is $R\leq\frac{C}{2}$. Next, we prove Theorem \ref{theorem:star} for the case where $d^{(t)}_{i,j}$ is fixed and equal to one, $\forall(i,j)\in\mathcal{E'}$ and $\forall{t}$.

\begin{IEEEproof}
{\it{Coding scheme:}} We choose two non-zero triplets $a=[a_1,a_2,a_3]$ and $b=[b_1,b_2,b_3]$ over the field $\mathbb{F}_{2^C}$ that satisfy
\begin{equation}\label{eq:star_coeff}
\begin{vmatrix}
    a_i& a_j\\
    b_i& b_j
\end{vmatrix}\neq{0}, \quad \forall{i,j}\in\{1,2,3\},i\neq{j}.
\end{equation}
Each source $i\in\mathcal{S}$ generates a transmission in $\mathbb{F}_{2^C}$ as follows
\begin{equation}\label{eq:star1}
X_i^{(t)}=k_iW_i^{(2\lfloor\frac{t}{2}\rfloor)}+k_jW_j^{(2\lfloor\frac{t-3}{2}\rfloor)}+k_lW_l^{(2\lfloor\frac{t-3}{2}\rfloor)},
\end{equation}
where $k_i={a_i}$ or $k_i={b_i}$, $\forall{i}\in\{1,2,3\}$, i.e., for each message $W_i^{(t)}$ that node $i$ generates, it transmits two transmissions (one with coefficients $k_i=a_i$ and the other with coefficients $k_i=b_i$). As such, we set the rate of each source node $i\in\mathcal{S}$ to $R_i=\frac{C}{2}$. The relay node, $4$, generates a transmission
\begin{equation}\label{eq:star2}
X_4^{(t)}=X_1^{(t-1)}+X_2^{(t-1)}+X_3^{(t-1)}.
\end{equation}
We now show that the coding scheme defined in (\ref{eq:star1}) and (\ref{eq:star2}) implies that
\begin{align}
X_4^{(t)}&=X_1^{(t-1)}+X_2^{(t-1)}+X_3^{(t-1)}\nonumber\\
&=k_1W_1^{(2\lfloor{\frac{t-4}{2}}\rfloor)}+k_1W_1^{(2\lfloor{\frac{t-4}{2}}\rfloor)}+k_2W_2^{(2\lfloor{\frac{t-4}{2}}\rfloor)}\nonumber\\
&\quad +k_2W_2^{(2\lfloor{\frac{t-4}{2}}\rfloor)}+k_3W_3^{(2\lfloor{\frac{t-4}{2}}\rfloor)}+k_3W_3^{(2\lfloor{\frac{t-4}{2}}\rfloor)}\nonumber\\
&\quad +k_1W_1^{(2\lfloor\frac{t-1}{2}\rfloor)}+k_2W_2^{(2\lfloor\frac{t-1}{2}\rfloor)}+k_3W_3^{(2\lfloor\frac{t-1}{2}\rfloor)}\nonumber\\
&=k_1W_1^{(2\lfloor\frac{t-1}{2}\rfloor)}+k_2W_2^{(2\lfloor\frac{t-1}{2}\rfloor)}+k_3W_3^{(2\lfloor\frac{t-1}{2}\rfloor)}\label{eq:form_star}.
\end{align}

\par {\it{Decoding:}} Each decoder $i\in\mathcal{S}$ obtains two independent equations for each two time instants $t+1$ and $t+2$ and is able to decode two messages $W_j^{(2\lfloor\frac{t-1}{2}\rfloor)}$, $\forall{j}\in\mathcal{S}\setminus\{i\}$.

\par We verify that this NC scheme enables all source nodes to recover all the messages with a delay of $L+1$, where $L=2$, using induction over $t$. At time $t=0$ the claim is trivial and, by the inductive assumption, up to time $t$ node $i$ can recover $\mathcal{W}_i^{2\lfloor{\frac{t}{2}}\rfloor},\mathcal{W}_j^{2\lfloor\frac{t-3}{2}\rfloor}$, $\forall{j}\in\mathcal{S}\setminus\{i\}$. At time $t+1$ and $t+2$, node $i$ obtains two transmissions, $X_4^{(t)}$ and $X_4^{(t+1)}$, respectively, and since the coefficients $a$ and $b$ were chosen according to (\ref{eq:star_coeff}), it is able to decode two messages, $W_j^{(2\lfloor\frac{t-1}{2}\rfloor)}$, $j\in\mathcal{S}\setminus\{i\}$, thereby proving the claim. Moreover, this claim also proves that this NC scheme is a RT coding scheme since the maximum delay is $L+t+1$ for all $t$, where $L=2$.
\end{IEEEproof}
Now we present the proof of Theorem \ref{theorem:star} for the case where $d^{(t)}_{i,j}$ is an arbitrary integer that can be different for each transmission over the edge $(i,j)\in\mathcal{E'}$ but is bounded by $D$.

\begin{figure}[!h]
\centering
    \psfrag{a}[][][1]{$p$}
    \psfrag{b}[][][1]{$q$}
    \psfrag{c}[][][1]{$Data$}
    \psfrag{e}[][][1]{$u$}
    \psfrag{d}[][][1]{$k$}
    \includegraphics[width=2.5in,height=0.2in]{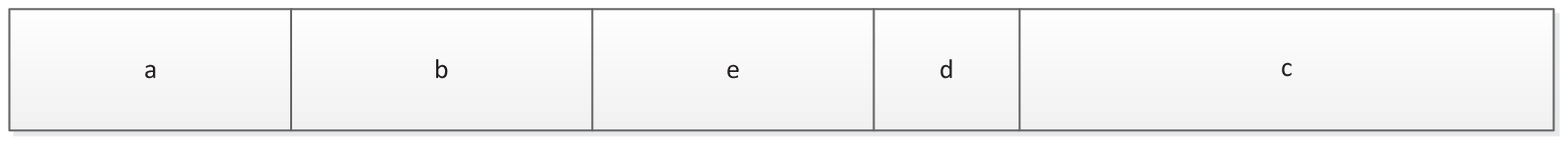}
    \caption{Packet header of an unsynchronized coding scheme for a star topology.}
    \label{fig:packet_star}
\end{figure}

\begin{IEEEproof}
We first note that Equation (\ref{eq:star2}) is no longer applicable for the unsynchronized case, and therefore, we introduce a new coding scheme for this case.
\par{\it{Coding scheme:}} The relay node, $4$, transmits a linear combination in $\mathbb{F}_{2^C}$, as follows
\begin{equation}\label{eq:star_random}
X_{4}^{(t)}=k_1W_1^{(p)}+k_2W_2^{(q)}+k_3W_3^{(u)},
\end{equation}
where $k_i={a_i}$ or $k_i={b_i}$, $\forall{i}\in\{1,2,3\}$, and $W_1^{(p)},W_2^{(q)},W_3^{(u)}$ are some arbitrary messages from the sets $\mathcal{W}_1^t, \mathcal{W}_2^t$ and $\mathcal{W}_3^t$, respectively. Namely, relay node $4$ transmits two independent transmissions (one with coefficients $k_i=a_i$ and the other with coefficients $k_i=b_i$). Similarly, each source node $i\in\mathcal{S}$ transmits
\begin{equation}
X_i^{(t)}=k_iW_i^{(t)}+k_jW_j^{(p)}+k_lW_l^{(q)},
\end{equation}
where $j,l\in\mathcal{S}\setminus\{i\}$.
\par {\it{Decoding:}} Each decoder $i\in\mathcal{S}$ obtains two independent equations and is able to decode two messages $W_j^{(p)},W_l^{(q)}$, $j,l\in\mathcal{S}\setminus\{i\}$, and $W_j^{(p)},W_l^{(q)}$ are some arbitrary messages from the sets $\mathcal{W}_j^t, \mathcal{W}_l^t$, respectively.

\par We verify that the relay node, $4$, is able to produce a transmission in the form of (\ref{eq:star_random}) at any particular time instant $t$. First, we add a header to each transmission to indicate which messages from the set $\mathcal{W}_i^t$ were encoded, $\forall{i}\in\mathcal{S}$ and another bit to indicate the coefficients ($k_i=a_i$ or $k_i=b_i$), as illustrated in Fig. \ref{fig:packet_star}. Second, we demand that relay node $4$ will decode and store the messages. We argue that the set of messages $\mathcal{W}_i^{t}$ held by source node $j$ is a subset of the messages $\mathcal{W}_i^{t'}$ held by node $4$ for all $t$, $\forall{i}\in\mathcal{S}\setminus\{j\}$. We verify the claim using induction. At time $t=0$ the claim is trivial, and therefore, we inductively assume that up to time $t$ the claim is true. We demonstrate the claim subsequently, where source node $j$ receives two transmissions, $k_1W_1^{(p)}+k_2W_2^{(q)}+k_3W_3^{(u)}$, one with coefficients $a_i$ and the other with coefficients $b_i$, $\forall{i}\in\{1,2,3\}$. Since by the inductive assumption node $j$ has messages $\mathcal{W}_j^{2\lfloor{\frac{t}{2}}\rfloor}$ and the coefficients were chosen according to (\ref{eq:star_coeff}), the node is able to decode two messages and the claim is established. To conclude, each source node $l\in\mathcal{S}$ is able to decode the messages sent by $\mathcal{S}\setminus\{l\}$ at any particular time instant $t$ independent of $d^{(t)}_{i,j}$, $\forall(i,j)\in\mathcal{E'}$. Furthermore, the maximum delay is bounded by $DL+t+1$ for all $t$, where $L=2$, and therefore this NC scheme is a RT coding scheme.
\end{IEEEproof}

\par In the corresponding wireless model, i.e., a star network $\mathcal{G(V,E)}$, applying our coding scheme requires two transmissions by the relay node for each triplet of messages, $W_1^{(t)},W_2^{(t)},W_3^{(t)}$, where using a simple routing scheme yields three transmissions. Hence, by exploiting the broadcast ability of the wireless medium, we obtain an energy efficient coding scheme for the wireless setting.

\subsection{Line-Star Topology}\label{sec:line_star}
\begin{figure}[!h]
\centering
    \psfrag{a}[][][1]{$1$}
    \psfrag{b}[][][1]{$2$}
    \psfrag{c}[][][1]{$3$}
    \psfrag{d}[][][1]{$4$}
    \psfrag{e}[][][1]{$4'$}
    \psfrag{f}[][][1]{$5$}
    \psfrag{g}[][][1]{$5'$}
    \includegraphics[width=2.5in]{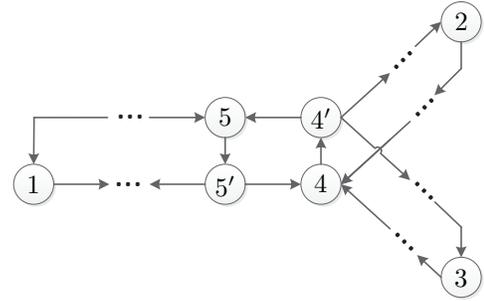}
    \caption{Graph $\mathcal{G'(V',E')}$ of a line-star topology, where $\mathcal{S}=\{1,2,3\}$. A line topology is formed between nodes $1$ and $5$, and a star topology is formed around node $4$.}
    \label{fig:star_line}
\end{figure}

The line-star topology is generally defined as a combination of line and star topologies (Fig. \ref{fig:star_line}), in which a line of nodes connects a source node to the star topology structure. The upper bound of the equal rate, which is identical to that for the star topology in Section \ref{sec:star}, is proved in the same manner. We demonstrate the coding scheme for this topology in the following section.

\par First, we show a  case where $d^{(t)}_{i,j}$ is fixed and equal to one, $\forall(i,j)\in\mathcal{E'}$ and $\forall{t}$ and we consider another single relay node, $5$, between nodes $1$ and $4$ (Fig. \ref{fig:star_line}).
\par {\it{Coding scheme:}}
Relay node $4$ and each source node $i\in\mathcal{S}$ perform coding in a manner identical to that described for the star topology in Section \ref{sec:star}. Node $5$ generates transmissions in the form of (\ref{eq:form_star}) in the same manner as the relays of the line topology in Section \ref{sec:line}, i.e.,
\begin{equation}
X_5^{(t)}=X_{1}^{(t-1)}+X_{4}^{(t-1)}+X_{5}^{(t-2)}\label{eq:form_star_mod}.
\end{equation}
This coding implies that
\begin{align}
X_5^{(t)}&=X_{1}^{(t-1)}+X_{4}^{(t-1)}+X_{5}^{(t-2)}\nonumber\\
&=k_1W_1^{(2\lfloor{\frac{t-3}{2}}\rfloor)}+k_1W_1^{(2\lfloor{\frac{t-3}{2}}\rfloor)}+k_1W_1^{(2\lfloor{\frac{t-1}{2}}\rfloor)}\nonumber\\
&\quad +k_2W_2^{(2\lfloor{\frac{t-4}{2}}\rfloor)}+k_2W_2^{(2\lfloor{\frac{t-4}{2}}\rfloor)}+k_2W_2^{(2\lfloor{\frac{t-2}{2}}\rfloor)}\nonumber\\
&\quad +k_3W_3^{(2\lfloor{\frac{t-4}{2}}\rfloor)}+k_3W_3^{(2\lfloor{\frac{t-4}{2}}\rfloor)}+k_3W_3^{(2\lfloor{\frac{t-2}{2}}\rfloor)}\nonumber\\
&=k_1W_1^{(2\lfloor\frac{t-1}{2}\rfloor)}+k_2W_2^{(2\lfloor\frac{t-2}{2}\rfloor)}+k_3W_3^{(2\lfloor\frac{t-2}{2}\rfloor)}\label{eq:form_line_star}.
\end{align}

\par {\it{Decoding:}} The decoding process is the same as the coding scheme of the star topology in Section \ref{sec:star}.

\par We verify that this NC scheme enables all source nodes to recover all the messages with a delay of $L+1$, where $L=3$, using induction over $t$. At time $t=0$ the claim is trivial and, by the inductive assumption, up to time $t$ node $i$ can recover $\mathcal{W}_i^{2\lfloor{\frac{t}{2}}\rfloor},\mathcal{W}_j^{2\lfloor\frac{t-3}{2}\rfloor}$, $\forall{j}\in\mathcal{S}\setminus\{i\}$. At time $t+1$ and $t+2$, node $i$ obtains two transmissions, $X_r^{(t)}$ and $X_r^{(t+1)}$, respectively, where $r\in\mathcal{V}\setminus\mathcal{S}$, and since the coefficients $a$ and $b$ were chosen according to (\ref{eq:star_coeff}), it is able to decode two messages, $W_j^{(2\lfloor\frac{t-1}{2}\rfloor)}$, $j\in\mathcal{S}\setminus\{i\}$, thereby proving the claim. Moreover, this claim also proves that this NC scheme is a RT coding scheme since the maximum delay is $L+t+1$ for all $t$, where $L=3$.

\par Now we present the case where $d^{(t)}_{i,j}$ is an arbitrary integer that can be different for each transmission over the edge $(i,j)\in\mathcal{E'}$ but that is bounded by $D$.

\par We first note that Equation (\ref{eq:form_star_mod}) is no longer applicable for the unsynchronized case, and therefore, we introduce a new coding scheme for this case.
\par{\it{Coding scheme:}} Relay node $4$ and each source node $i\in\mathcal{S}$ perform coding in a manner identical to that described for the unsynchronized case of the star topology in Section \ref{sec:star}. Node $5$ transmits
\begin{equation}
X_{5}^{(t)}=k_1W_1^{(p)}+k_2W_2^{(q)}+k_3W_3^{(u)},
\end{equation}
where $k_i={a_i}$ or $k_i={b_i}$, $\forall{i}\in\{1,2,3\}$, $W_1^{(p)},W_2^{(q)}$ and $W_3^{(u)}$ are some arbitrary messages from the sets $\mathcal{W}_1^t, \mathcal{W}_2^t$ and $\mathcal{W}_3^t$, respectively. Namely, it transmits two independent transmissions (one with coefficients $k_i=a_i$ and the other with coefficients $k_i=b_i$) with the same header as described in Fig. \ref{fig:packet_star}. Node $5$ needs to decode
all messages, $\mathcal{W}_1^{t}$, $\mathcal{W}_2^t$ and $\mathcal{W}_3^t$, and then to store them for this coding scheme to work. We argue that the set of messages $\mathcal{W}_i^{t}$ held by source node $j$ is a subset of the messages $\mathcal{W}_i^{t'}$ held by nodes $4$ and $5$ for all $t$, $\forall{i}\in\mathcal{S}\setminus\{j\}$. We verify the claim using induction. At time $t=0$ the claim is trivial, and therefore, we inductively assume that up to time $t$ the claim is true. We demonstrate the claim subsequently, where source node $j$ receives two transmissions, $k_1W_1^{(p)}+k_2W_2^{(q)}+k_3W_3^{(u)}$, one with coefficients $a_i$ and the other with coefficients $b_i$, $\forall{i}\in\{1,2,3\}$. Since by the inductive assumption node $j$ has messages $\mathcal{W}_j^{2\lfloor{\frac{t}{2}}\rfloor}$ and the coefficients were chosen according to (\ref{eq:star_coeff}), the node is able to decode two messages and the claim is established. To conclude, each source node $l\in\mathcal{S}$ is able to decode the messages sent by $\mathcal{S}\setminus\{l\}$ at any particular time instant $t$ independent of $d^{(t)}_{i,j}$, $\forall(i,j)\in\mathcal{E'}$. Furthermore, the maximum delay is bounded by $DL+t+1$ for all $t$, and therefore, this NC scheme is a RT coding scheme.

\par This process of adding a relay node to the star topology can now be extended to any given number of nodes connecting each source node as a line topology to the star topology structure. In the corresponding wireless model, i.e., a line-star network $\mathcal{G(V,E)}$, applying our coding scheme requires $2(|\mathcal{V}|-3)$ transmissions for each triplet of messages, $W_1^{(t)},W_2^{(t)},W_3^{(t)}$, i.e., two transmissions by each relay node, $\mathcal{V}\setminus\mathcal{S}$, where using a simple routing scheme yields $3(|\mathcal{V}|-3)$ transmissions. Hence, by exploiting the broadcast ability of the wireless medium, we obtain an energy efficient coding scheme for the wireless setting.

\par The topologies of line, star, and line-star are all from the same family in the sense that they all represent networks with a minimum cut of one between each two source nodes. Additionally, by applying a simple routing scheme to the wireless model of each of the topologies, we can achieve the maximum rate. However, by using NC, we achieve the minimum number of transmissions, i.e., an efficient energy consumption, without rate redundancy. The coding schemes presented in Sections \ref{sec:line}, \ref{sec:star} and \ref{sec:line_star} hold the properties of RT and are \emph{innovative} \cite{cite:inovite}. A coding scheme is called innovative if each incoming transmission to source node $i\in\mathcal{S}$ is not contained in the span of messages previously received by $i$. Furthermore, the line topology coding scheme of Section \ref{sec:line} also holds a property of \emph{instantly decodable}, a coding scheme in which a new message is decoded for each incoming transmission, i.e., a RT coding scheme with $c=0$. In the rest of the paper, we use these coding schemes as building blocks and exploit their properties to construct a coding scheme for a general network.

\section{Multicast Network}\label{sec:multicast}
In this section, we combine the line topology from Section \ref{sec:line} and the line-star topology from Section \ref{sec:line_star} and use them as building blocks to present a new coding scheme for a general multicast network. A general network is defined by a graph $\mathcal{G'(V',E')}$ with a set of source nodes $\mathcal{S}=\{1,2,3\}$ (Fig. \ref{fig:model_multicast}). Every network is shown to be decomposable into line-star and ring topologies, where a ring is defined as three special line topologies. The special line topology is valid for a ring if deleting that line will not decrease the minimal cut of the remaining source node, e.g., if deleting $P_{1;2}$ will not decrease $\mathcal{C}_{3;1,2}$. Each network may have many decompositions of those building blocks. However, we prove that there exists at least one decomposition that, using the coding schemes of the building blocks, achieves the equal rate capacity for a general wireless network, in which three source nodes communicate bidirectionally in a multicast manner.

\subsection{Capacity and Coding for a Multicast Network Based on Line and Star Topologies}
We denote the minimum of all the cuts that separate one source from the network by
\begin{equation}\label{eq:multicast_def_h}
h=\frac{\min_{i\in\mathcal{S}}\mathcal{C}_{i;\mathcal{S}\setminus\{i\}}}{C}.
\end{equation}
Our main result is summarized in the following theorem.

\begin{theorem}\label{theorem:model_multicast}
For any network $\mathcal{G'(V',E')}$ with an arbitrary delay bounded by $D$, there exists a RT coding scheme that achieves the equal rate capacity, which is $\frac{hC}{2}$. Furthermore, the coding scheme includes a fixed header per transmission of $3\lceil\log_{2}2D\rceil+1+\lceil\log_2{h}\rceil$ bits.
\end{theorem}

The upper bound is obtained by the standard minimum-cut arguments \cite{cite:ford}. We assume, without loss of generality, that $h=\frac{\mathcal{C}_{i;\mathcal{S}\setminus\{i\}}}{C}$ for some $i\in\mathcal{S}$. Therefore, we get an upper bound of $R_j+R_l\leq{hC}$, $j,l\in\mathcal{S}\setminus\{i\}$ and the equal rate upper bound is $R\leq\frac{hC}{2}$. This means that the three users can exchange information that was generated up to time $t$ at rate $R\leq\frac{hC}{2}$ and with delay at most $LD+t+1$, $\forall{t}$.

\begin{figure*}[bp]
\centering
    \subfloat[Network model\label{fig:multicast_example_case1}]{
    \psfrag{a}[][][0.8]{$3$}
    \psfrag{b}[][][0.8]{$2$}
    \psfrag{c}[][][0.8]{$1$}
    \psfrag{d}[][][0.8]{$W^{(t)}_{2}$}
    \psfrag{e}[][][0.8]{$W^{(t)}_{1}$}
    \psfrag{f}[][][0.8]{$W^{(t)}_{3}$}
    \psfrag{r}[][][0.8]{$4$}
    \psfrag{t}[][][0.8]{$5$}
    \psfrag{y}[][][0.8]{$6$}
    \psfrag{u}[][][0.8]{$7$}
    \psfrag{v}[][][0.8]{$8$}
    \psfrag{o}[][][0.8]{$9$}
    \psfrag{p}[][][0.8]{$10$}
    \psfrag{x}[][][0.8]{$11$}
    \includegraphics[scale=.35]{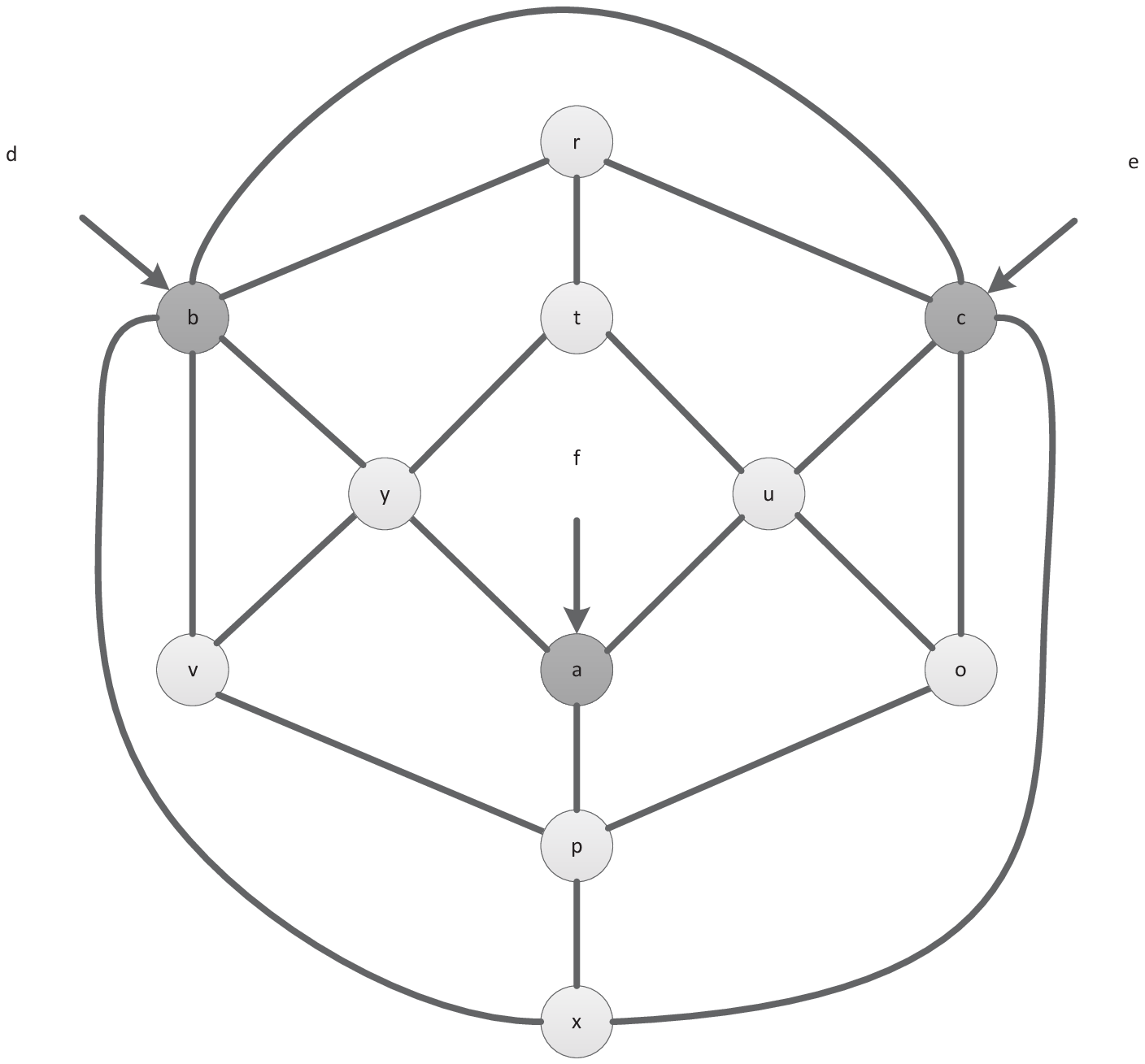}%
    }
    \hfil
    \subfloat[Ring topology - $\mathcal{R}$\label{fig:multicast_example_case2}]{
    \psfrag{a}[][][0.8]{$3$}
    \psfrag{b}[][][0.8]{$2$}
    \psfrag{c}[][][0.8]{$1$}
    \psfrag{d}[][][0.8]{$W^{(t)}_{2}$}
    \psfrag{e}[][][0.8]{$W^{(t)}_{1}$}
    \psfrag{f}[][][0.8]{$W^{(t)}_{3}$}
    \psfrag{r}[][][0.8]{$4$}
    \psfrag{t}[][][0.8]{$5$}
    \psfrag{y}[][][0.8]{$6$}
    \psfrag{u}[][][0.8]{$7$}
    \psfrag{v}[][][0.8]{$8$}
    \psfrag{o}[][][0.8]{$9$}
    \psfrag{p}[][][0.8]{$10$}
    \psfrag{x}[][][0.8]{$11$}
    \includegraphics[scale=.35]{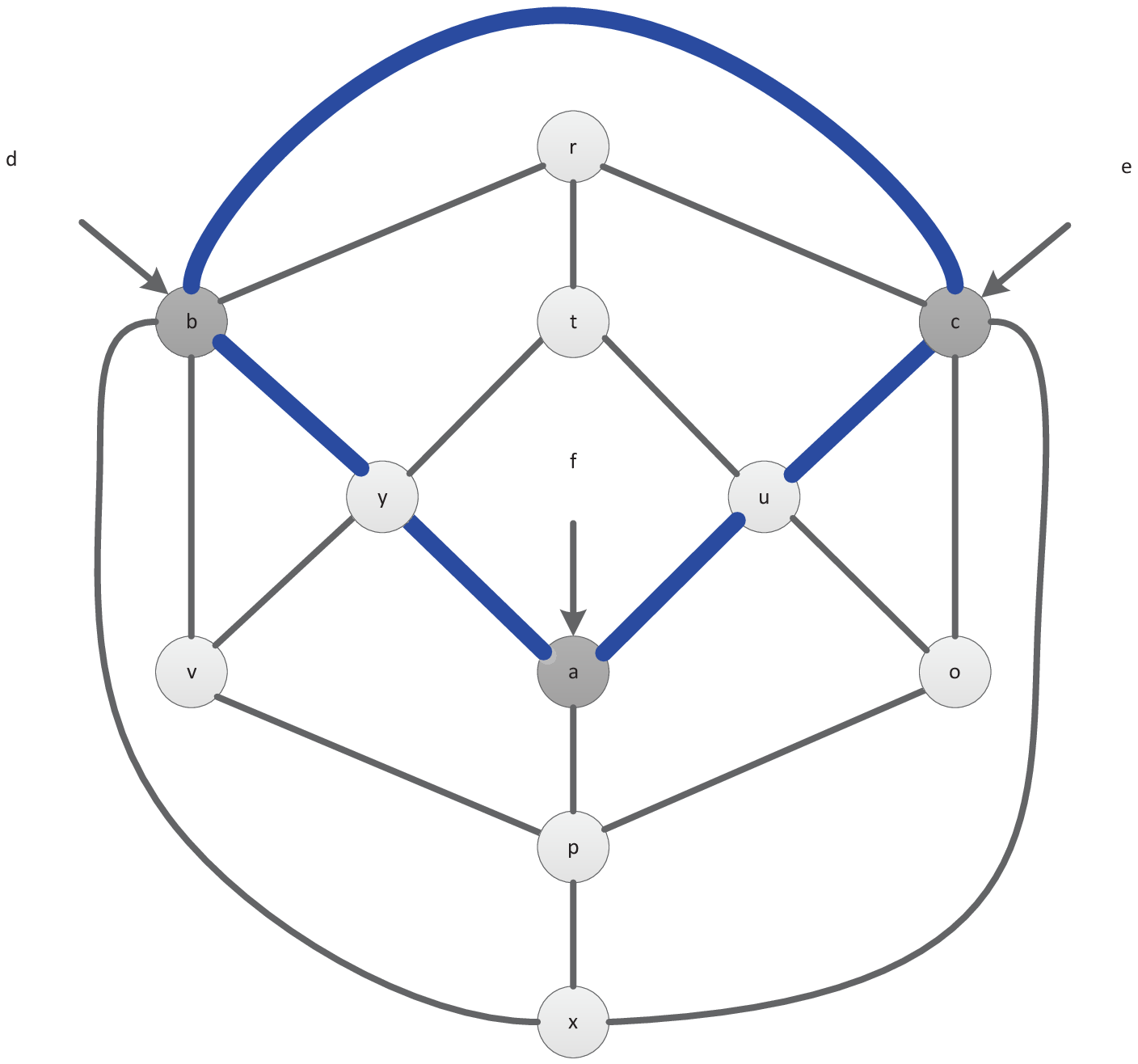}%
    }
    \hfil
    \subfloat[Line-star topology - $\mathcal{Q}$\label{fig:mlticast_example_case3}]{
    \psfrag{a}[][][0.8]{$3$}
    \psfrag{b}[][][0.8]{$2$}
    \psfrag{c}[][][0.8]{$1$}
    \psfrag{d}[][][0.8]{$W^{(t)}_{2}$}
    \psfrag{e}[][][0.8]{$W^{(t)}_{1}$}
    \psfrag{f}[][][0.8]{$W^{(t)}_{3}$}
    \psfrag{r}[][][0.8]{$4$}
    \psfrag{t}[][][0.8]{$5$}
    \psfrag{y}[][][0.8]{$6$}
    \psfrag{u}[][][0.8]{$7$}
    \psfrag{v}[][][0.8]{$8$}
    \psfrag{o}[][][0.8]{$9$}
    \psfrag{p}[][][0.8]{$10$}
    \psfrag{x}[][][0.8]{$11$}
    \includegraphics[scale=.35]{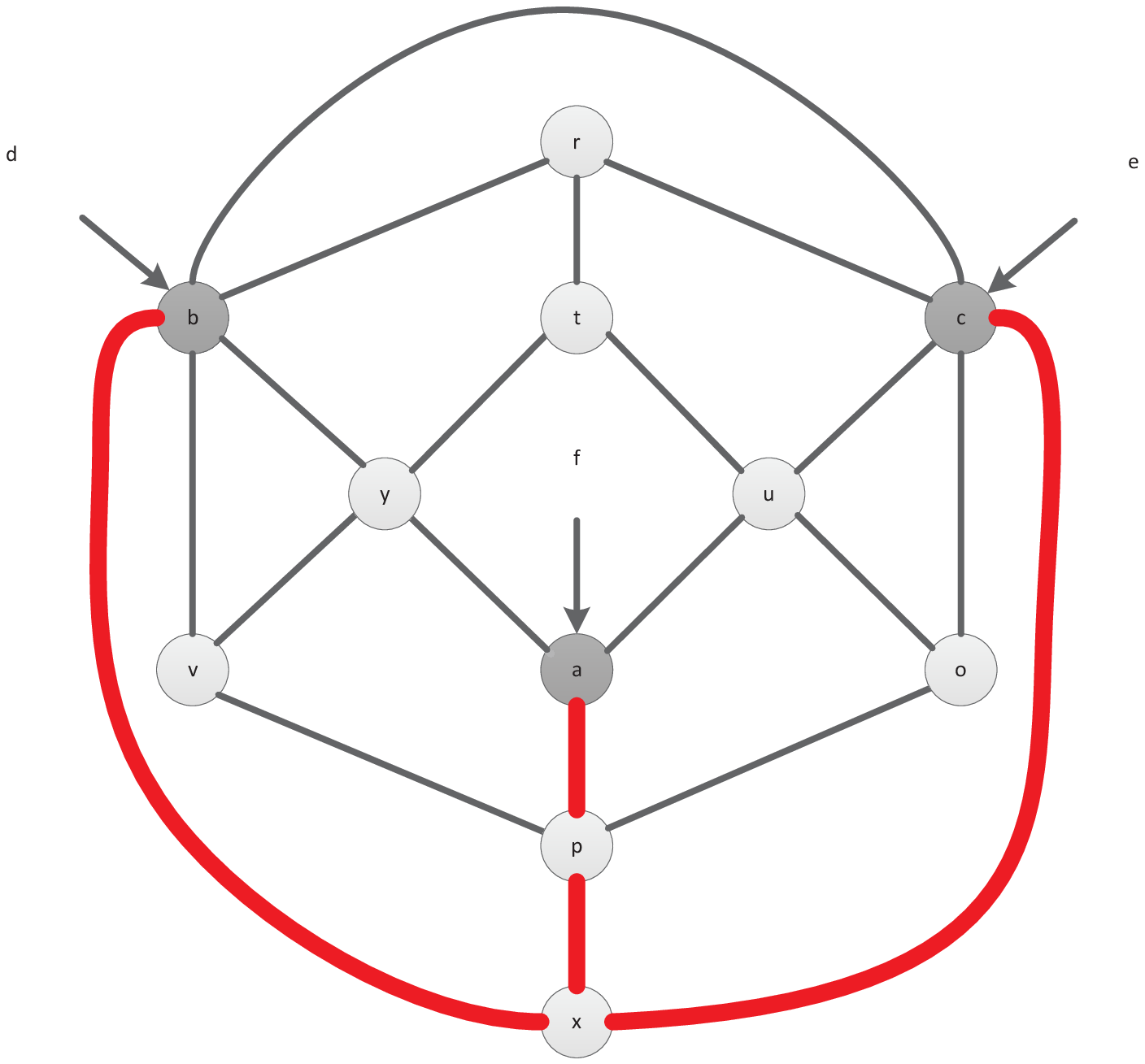}%
    }
    \caption{A network $\mathcal{G}$ is shown in \protect\subref{fig:multicast_example_case1}, where all the source nodes $1,2$ and $3$ communicate in a bidirectional multicast manner. By using the ring and line-star building blocks approach from Section \protect\ref{sec:multicast}, we get the set $\mathcal{R}$, which represents the ring topologies in the network, illustrated in \protect\subref{fig:multicast_example_case2}, and the set $\mathcal{Q}$, which represents the line-star topologies, as illustrated in \protect\subref{fig:mlticast_example_case3}.}
    \label{fig:fullExample_multicast}
\end{figure*}

\par To prove the achievability of Theorem \ref{theorem:model_multicast}, we first introduce Lemma \ref{lemma:disjoint_multicast}, which shows how to partition each network $\mathcal{G'}$ into sub-topologies of line and line-star networks. The line topologies are formed by a set of rings $\mathcal{R}$, where each $r\in\mathcal{R}$ is defined in the following definition.
\begin{definition}[Ring]
A ring, $r$, is defined for a graph, $\mathcal{G'(V',E')}$, with three source nodes, $\mathcal{S}=\{1,2,3\}$, as a set of edges, i.e., $r\subseteq\mathcal{E'}$, which forms three bidirectional paths between each two source nodes under the condition that removing a path between $i$ and $j$ is not lessening the minimum cut of the remaining source node $l$, i.e., $\mathcal{C}_{l;i,j}$, $i,j,l\in\mathcal{S}$.
\end{definition}

\begin{figure}[h!]
\centering
    \subfloat[Ring topology\label{fig:example_multicast1}]{%
        \psfrag{a}[][][1]{$1$}
        \psfrag{b}[][][1]{$2$}
        \psfrag{c}[][][1]{$3$}
        \includegraphics[scale=0.6]{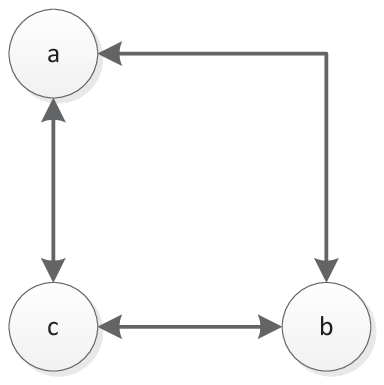}%
    }
    \quad\quad\quad\quad
    \subfloat[Line-star topology\label{fig:example_multicast2}]{%
        \psfrag{a}[][][1]{$1$}
        \psfrag{b}[][][1]{$2$}
        \psfrag{c}[][][1]{$3$}
        \includegraphics[scale=0.6]{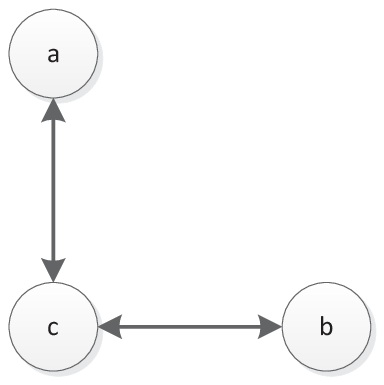}%
    }
    \caption{Schematic illustrations of \protect\subref{fig:example_multicast1} an element in $\mathcal{R}$ (a ring) and \protect\subref{fig:example_multicast2} an element in $\mathcal{Q}$ (a line-star topology).}
    \label{fig:example_multicast}
\end{figure}

Each ring, $r$, in a graph contributes a rate of $C$ to each source node by using the line topology coding scheme of Section \ref{sec:line} at each bidirectional path, i.e., $r_1\bigcap{r_2}=\emptyset$, $\forall{r_1},r_2\in\mathcal{R}$. For example, consider the ring in Fig. \ref{fig:example_multicast1}. The line-star topologies are defined by a set $\mathcal{Q}$, where each $q\in\mathcal{Q}$ is defined by a union two bidirectional paths, where both  paths leave the same source node, but each path is destined for another node, as illustrated in Fig. \ref{fig:example_multicast2}.

\par Each ring in also an element in $\mathcal{Q}$. However, each $q\in\mathcal{Q}$ contributes only a rate of $\frac{C}{2}$ to each source node by using the line-star topology coding scheme from Section \ref{sec:line_star}, i.e., $q_1\bigcap{q_2}=\emptyset$, $\forall{q_1,q_2}\in\mathcal{Q}$. Although there exist many decompositions of $\mathcal{R}$ and $\mathcal{Q}$ in a network $\mathcal{G'}$, in the following lemma we show that by first finding the maximum number of rings, we can assure that there exist enough $\mathcal{R}$ and $\mathcal{Q}$ to achieve the equal rate upper bound.

\begin{lemma}\label{lemma:disjoint_multicast}
For a network $\mathcal{G'(V',E')}$, there exist $\mathcal{R}$ and $\mathcal{Q}$ such that $|\mathcal{R}|+\frac{|\mathcal{Q}|}{2}\geq\frac{h}{2}$, where $\mathcal{R}\cap\mathcal{Q}=\emptyset$. Namely, $\mathcal{R}$ and $\mathcal{Q}$ have no mutual edges.
\end{lemma}

\begin{IEEEproof}[Proof of Lemma \ref{lemma:disjoint_multicast}]
The proof is by construction. First, we search for the maximum number of rings, $|\mathcal{R}|$. Then, we construct a network $\mathcal{G''}$ that is the network $\mathcal{G'}$ without $\mathcal{R}$, i.e., we remove the edges in $\mathcal{R}$ from $\mathcal{G'}$. Without loss of generality, we assume that between source nodes $1$ and $2$ there are no more paths, such that deleting them from $\mathcal{G''}$ will not decrease $\mathcal{C}_{3;1,2}$. We can always find a pair of source nodes that satisfies this condition, because otherwise, we could increase $\mathcal{R}$. Finally, we find a set of new paths $\mathcal{Q}$ in $\mathcal{G''}$, where $|\mathcal{Q}|=\frac{\mathcal{C''}_{1;2}}{C}$. Each path, $P_{1;2}\in\mathcal{P}_{1;2}$, from node $1$ to node $2$ shares at least one common node with a special path, $P_{1;3}\in\mathcal{P}_{1;3}$ from node $1$ to node $3$. This path, $P_{1;3}$, shares no common nodes with all the other paths, $\mathcal{P}_{1;2}\setminus\{P_{1;2}\}$, between nodes $1$ and $2$. At least one special path $P_{1;3}$ exists that corresponds to each selection of $P_{1;2}$, since otherwise deleting $P_{1;2}$ will not decrease $\mathcal{C}_{3;1,2}$. As a result, finding the path $P_{1;2}$ and one of the special paths $P_{1;3}$ that corresponds to it is equivalent to finding a line-star topology in the network. The line-star topology consists of the union between $P_{1;2}$ and $P_{1;3}$, i.e., the union of $P_{1;2}$ and $P_{1;3}$ is a $q\in\mathcal{Q}$. Therefore,

\begin{align}
|\mathcal{Q}|&\stackrel{(a)}{=}\frac{\mathcal{C''}_{1;2}}{C}\\
&\stackrel{(b)}{=}\frac{\min\{\mathcal{C''}_{1;2,3},\mathcal{C''}_{2;1,3}\}}{C}\\
&\stackrel{(c)}{=}\frac{\min\{\mathcal{C}_{1;2,3},\mathcal{C}_{2;1,3}\}}{C}-2|\mathcal{R}|\\
&\stackrel{(d)}{\geq}{h}-2|\mathcal{R}|,
\end{align}
where $(a)$ follows from the fact that there are $\frac{\mathcal{C''}_{1;2}}{C}$ paths $P_{1;2}$ that has a corresponding path $P_{1;3}$ that creates a line-star topology in the network. $(b)$ is true since node $3$ can only be in one of the cuts separating nodes $1$ and $2$, and $\mathcal{C}_{1;2}$ is the minimization of all the cuts separating them. $(c)$ is the transition to the network $\mathcal{G'}$ and $d$ follows from (\ref{eq:multicast_def_h}).
\end{IEEEproof}
Next, we provide the proof for Theorem \ref{theorem:model_multicast}.

\begin{IEEEproof}[Proof of Theorem \ref{theorem:model_multicast}]
Using the line topology coding scheme from Section \ref{sec:line} at each path in a ring $r\in\mathcal{R}$ yields a rate of $R_i=|\mathcal{R}|C$, $\forall{i}\in\mathcal{S}$. Furthermore, using the line-star topology of Section \ref{sec:line_star} at each $q\in\mathcal{Q}$ yields a rate of $R_i=\frac{|\mathcal{Q}|C}{2}$, $\forall{i}\in\mathcal{S}$. Therefore, using Lemma \ref{lemma:disjoint_multicast} and the fact that $\mathcal{R}\cap\mathcal{Q}=\emptyset$, we obtain that $R\geq\frac{hC}{2}$, and since this is an upper bound, we have an equality. Moreover, we obtain an asynchronous coding scheme with a maximum delay of $LD+t+1$ for all $t$, by using a header of $3\lceil\log_{2}2D\rceil+1$ bits according to the coding scheme of the star topology from Section \ref{sec:star} and another $\lceil\log_2{h}\rceil$ bits are used to form disjoint line and line-star topologies in the network. Therefore, we obtain a RT NC scheme that achieves the equal rate capacity.
\end{IEEEproof}

\subsection{Example}
In this subsection, we show a multicast network in Fig. \ref{fig:fullExample_multicast}. This network is decomposed into ring and line-star topologies, and then we use the coding schemes of those two canonical topologies to obtain a new coding scheme that achieves the equal rate capacity. In Fig. \ref{fig:multicast_example_case2}, we visualize the ring topologies in the network. In this example, only one element in $\mathcal{R}$ exists, since there is only one path from node $1$ to node $3$ that deleting it from the network will not reduce the minimum cut of node $2$, i.e., $\mathcal{C}_{2;1,3}$. Note that this is not the only choice for a ring in the network, e.g., a different choice, $\mathcal{R'}$, include nodes $1,2,3,6,7$ and $11$ and all the direct links that connect them. As a negative example, consider a path from node $1$ to node $3$ through nodes $9$ and $10$. This path cannot be a part of a ring topology since deleting it will decrease the minimum cut of node $2$, i.e., $\mathcal{C}_{2;1,3}$.

\par In Fig. \ref{fig:mlticast_example_case3}, we present the line-star topologies in the network. In case $\mathcal{R'}$ is chosen a different choice of a line-star topology include nodes $1,2,3,8,9$ and $10$ and all the direct links that connect them. By using those topologies as building blocks, we achieve a rate of $R=1.5C$, which is also the equal rate upper bound of this network, i.e., $\frac{h}{2}$, where $h=3C$.

\subsection{Algorithm to Find the Building Blocks}\label{sec:algorithm_multicast}
The difficulty of finding $\mathcal{R}$ and $\mathcal{Q}$, i.e., the sets of edges that serve as building blocks of ring and line-star topologies in the network, is shown to be equivalent to solving a binary multicommodity flow problem. This class of optimization problem was shown to be a NP-complete problem \cite{cite:karp}. However, there exist many approximation algorithms for the integral multicommodity flow problem \cite{cite:approx,cite:approx2}. Furthermore, finding $\mathcal{R}$ and $\mathcal{Q}$ can be facilitated by introducing the upper bound $|\mathcal{R}|+|\mathcal{Q}|\leq\min_{i\in\mathcal{S}}deg(i)$,
where $deg(i)$ is defined to be the degree of node $i\in\mathcal{V'}$, i.e., the number of edges that are initiated at node $i$. The two sets $\mathcal{R}$ and $\mathcal{Q}$ represent disjoint edges and each element in these sets represents a path that passes through each source node and therefore, we obtain an upper bound of the size of the sets.

\par To find the set $\mathcal{R}$, we introduce an optimization problem. We denote by $\mathcal{O}_j$ and $\mathcal{I}_j$ the output and input flows from node $j$, respectively. Furthermore, the flow, denoted by $\mathbf{f}$, is a binary vector that represents the flow in each edge, where an element $f_{(i,j)}\in\mathbf{f}$ represents a binary flow in edge $(i,j)\in\mathcal{E'}$.

\par In this problem, we would like to find the maximum number of disjoint paths from node $i$ to node $j$ that deleting them from the network will not decrease the maximum number of disjoint paths from nodes $i$ and $j$ to node $l$, $\mathcal{C}_{l;i,j}$, where $i,j,l\in\mathcal{S}$. Therefore, we demand maximal input flow to node $l$, i.e.,
\begin{equation}\label{eq:opt_con}
\mathcal{I}_l=\frac{\mathcal{C}_{l;i,j}}{C}.
\end{equation}
Additionally, we demand that $\mathcal{O}_l=0$, i.e., the entire flow will originate at nodes $i$ and $j$ and that $\mathcal{I}_i=0$ since we are interested in finding a flow that terminates at nodes $j$ and $l$. With condition (\ref{eq:opt_con}) satisfied, we maximize the flow that originates at node $i$ and terminates at node $j$, i.e., the flow that is consumed by node $j$, $\mathcal{I}_j-\mathcal{O}_j$. Since $\mathbf{f}$ is a binary flow and each relay, $m\in\mathcal{V'}\setminus\mathcal{S}$, has equal input and output flows, we get two sets of disjoint paths. First, paths that terminate at node $l$ and satisfy condition (\ref{eq:opt_con}), and second, paths from $i$ to $j$. Because the two sets are disjoint, even if we remove the edges associated with the second set, i.e., paths from $i$ to $j$, condition (\ref{eq:opt_con}) will still be satisfied.

\begin{equation}
\begin{aligned}\label{eq:multicast_findLine}
\underset{{\mathbf{f}}}{\text{maximize}} & \quad\mathcal{I}_j-\mathcal{O}_j \\
\text{subject to} & \quad\mathcal{O}_m = \sum_{n:(m,n)\in\mathcal{E'}}f_{(m,n)}, \; m = 1 \ldots, |\mathcal{V}| \\
& \quad\mathcal{I}_m = \sum_{n:(n,m)\in\mathcal{E'}}f_{(n,m)}, \; m = 1 \ldots, |\mathcal{V}|\\
& \quad\mathcal{I}_l = \frac{\mathcal{C}_{l;i,j}}{C} \\
& \quad\mathcal{I}_{i} = 0 \\
& \quad\mathcal{O}_l = 0 \\
& \quad\mathcal{O}_i+\mathcal{O}_j = \mathcal{I}_{j}+\mathcal{I}_{l} \\
& \quad\mathcal{O}_k = \mathcal{I}_k, \; k = 4, \ldots, |\mathcal{V}|,
\end{aligned}
\end{equation}
$\forall{i,j,l}\in\mathcal{S}$.
\begin{algorithm}[!h]
\caption{Find $\mathcal{R}$}\label{algo:multicast_find_line}
\begin{algorithmic}[1]
\STATE $\mathcal{R}\leftarrow\emptyset$
\WHILE{True}
    \STATE Solve problem (\ref{eq:multicast_findLine}) for each pair $i,j\in\mathcal{S}$.
    \STATE Find $\mathcal{I'}_j$ which is the input flow to node $j$ after removing the cyclic flow.
    \IF{$\mathcal{I'}_j={0}$ for any $j\in\mathcal{S}$}
        \STATE Break.
    \ENDIF
    \STATE For each pair $i,j\in\mathcal{S}$, choose randomly one path, $P_{i;j}$, out of $\mathbf{f}$ between nodes $i$ and $j$.
    \STATE Add the three paths $P_{i;j}$ for each pair $i,j\in\mathcal{S}$ to $\mathcal{R}$ and remove them from the network.
\ENDWHILE
\end{algorithmic}
\end{algorithm}
\par The result of applying this optimization problem is a binary flow that satisfies the conditions of the optimization problem. From this flow, we subtract the cyclic flow, which is defined as the flow that originates and terminates at the same node, i.e., paths from node $j$ to node $j$ that do not pass through nodes $i$ and $l$, $i,j,l\in\mathcal{S}$. After the substraction, we obtain a new input flow to node $j$, $\mathcal{I'}_j$. This flow, $\mathcal{I'}_j$, represents the maximum number of paths between nodes $i$ and $j$, where condition (\ref{eq:opt_con}) is satisfied even if we delete those paths from the network, i.e., we remove their edges from the graph. Therefore, they are applicable to the set $\mathcal{R}$. Algorithm \ref{algo:multicast_find_line} describes how to find all of the paths in $\mathcal{R}$.

\begin{algorithm}[!h]
\caption{Find $\mathcal{Q}$}\label{algo:multicast_find_star}
\begin{algorithmic}[1]
\STATE Remove the edges associated with $\mathcal{R}$ from the graph.
\STATE $\mathcal{Q}\leftarrow\emptyset$
\WHILE{True}
    \IF{$|\mathcal{Q}|\geq{h-2|\mathcal{R}|}$}
        \STATE Break.
    \ENDIF
    \STATE Find two source nodes, $i,j\in\mathcal{S}$, which have no path between them that deleting it from the network will not decrease $\mathcal{C}_{l;i,j}$, $l\in\mathcal{S}\setminus\{i,j\}$.
    \STATE Find a path between nodes $i$ and $j$, $P_{i;j}$, which deleting it from the network will minimally decrease $\mathcal{C}_{l;i,j}$.
    \STATE Remove the edges associated with $P_{i;j}$ from the graph.
    \STATE Find all the paths from node $l$ to node $i$ in the new network and delete them from the network, i.e., remove their edges from the graph.
    \STATE Recover the path $P_{i;j}$ to the network and find a path $P_{i;l}$ in the new network.
    \STATE Add the union of $P_{i;l}$ and $P_{i;j}$ to $\mathcal{Q}$ and remove it from the network, i.e., remove the edges associated with $\mathcal{Q}$ from $\mathcal{E'}$.
\ENDWHILE
\end{algorithmic}
\end{algorithm}

\par Next, we present Algorithm \ref{algo:multicast_find_star}, which describes how to find the line-star topologies, $\mathcal{Q}$. This algorithm is based on the assumption that we already deleted the set $\mathcal{R}$ from the network, i.e., we removed the edges associated with $\mathcal{R}$ from the graph. Therefore, there exists at least one pair of source nodes $i$ and $j$ that have no path between them that deleting it will not decrease $\mathcal{C}_{l;i,j}$. After finding $i$ and $j$, we search for a path $P_{i;j}\in\mathcal{P}_{i;j}$ that deleting it from the network will cause the smallest possible decrement to $\mathcal{C}_{l;i,j}$, because we would like to avoid crossing other paths. Since deleting $P_{i;j}$ results in a decrement to $\mathcal{C}_{l;i,j}$, there exists at least one special path, $P_{i;l}\in\mathcal{P}_{i;l}$, from node $i$ to node $l$ that does not intersect with any of the other paths, $\mathcal{P}_{i;j}\setminus\{P_{i;j}\}$ from $i$ to $j$, i.e., it only shares a common node with $P_{i;j}$. To find this special path, we delete all the remaining paths from node $i$ to node $l$. By now, we have deleted from the network the path $P_{i;j}$ and all the remaining paths from node $i$ to node $l$. Next, by returning $P_{i;j}$ to the network, we assure that there exists a path between $i$ and $l$, $P_{i;l}$, which is the special path. A line-star topology structure, which is the union of $P_{i;j}$ and $P_{i;l}$, is then found.

\par
\begin{figure}[h!]
\centering
    \psfrag{b}[][][1]{$1$}
    \psfrag{c}[][][1]{$2$}
    \psfrag{d}[][][1]{$3$}
    \psfrag{a}[][][1]{$4$}
    \psfrag{p}[][][1]{$5$}
    \includegraphics[scale=0.6]{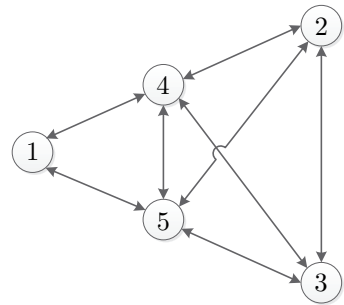}
    \caption{Graph $\mathcal{G(V,E)}$ with two star topologies, where $\mathcal{S}=\{1,2,3\}$.}
    \label{fig:multicast_algo_example}
\end{figure}

 To demonstrate Algorithm \ref{algo:multicast_find_star}, we apply it in an example. Graph $\mathcal{G(V,E)}$, which is illustrated in Fig. \ref{fig:multicast_algo_example}, contains three source nodes $1,2$ and $3$. We note that there is no path from node $1$ to node $2$ under the condition that deleting that path will not change the minimum cut of node $3$, i.e., $\mathcal{C}_{3;1,2}$. Therefore, there are no rings in the graph.
In the first step, we find a path between $1$ and $2$, $P_{1;2}$, that deleting it from the network decreases the minimum cut of node $3$, $\mathcal{C}_{3;1,2}$, by $C$, which is minimal, e.g., $P_{1;2}=\{(1,4),(4,2)\}$. Next, we delete $P_{1;2}$ from the network and then find all the remaining paths from node $3$ to node $1$ and delete them as well, e.g., $\{(3,5),(5,1)\}$. In the third phase, we return the path $P_{1;2}$ to the network and search for a path from node $1$ to node $3$, e.g., $P_{1;3}=\{(1,4),(4,3)\}$. Finally, we conclude that the union of $P_{1;2}$ and $P_{1;3}$ is an element in $\mathcal{Q}$, i.e., $q=\{(1,4),(2,4),(3,4)\}$, $q\in\mathcal{Q}$.

\section{Multiple Unicast Network}\label{sec:unicast}

In this section, we present the coding scheme for the multiple unicast network, depicted in Fig. \ref{fig:model_unicast}. This network is defined by a graph $\mathcal{G'(V',E')}$ with a set of source nodes $\mathcal{S}=\{1,2,3\}$ (Fig. \ref{fig:model_unicast}). In the multiple unicast network, each source node $i\in\mathcal{S}$ produces two different messages $W_{i\rightarrow{j}}^{(t)}$ and $W_{i\rightarrow{l}}^{(t)}$ that are intended for the two remaining source nodes $j,l\in\mathcal{S}\setminus\{i\}$, i.e., each two source nodes communicate bidirectionally in a unicast manner.

\subsection{Capacity and Coding for a Multiple Unicast Network Based on Line Topology}

Our coding scheme achieves the capacity region under an equal rate demand. Specifically, each two-way communication is carried out at the same rate, $R_{i\rightarrow{j}}=R_{j\rightarrow{i}}$, $\forall{i,j}\in\mathcal{S}$. To show the RT coding scheme for this network, we use the line topology coding scheme from Section \ref{sec:line}. Since each unicast session represents a flow from one source node to another, we show that no inter-flow coding is needed to achieve the capacity region. Under the equal rate assumption, we show that the minimum cut upper bound is achievable. However, in the multiple unicast general case the minimum cut upper bound is not tight \cite{cite:insufficiency}, i.e., our coding scheme is optimal under an equal rate assumption, but sub-optimal in the general case. Our main result is summarized in the following theorem.

\begin{theorem}\label{theorem:unicast}
For any network $\mathcal{G'(V',E')}$ with an arbitrary delay bounded by $D$, there exists a RT coding scheme with any set of rates within the equal rate capacity region, which is for all $i,j,l\in\mathcal{S}$,
\begin{equation}\label{eq:unicast_upper_bound}
\begin{aligned}
R_{i\rightarrow{j}}+R_{i\rightarrow{l}}&\leq{\mathcal{C}_{i;j,l}},\\
R_{i\rightarrow{j}}&\leq{\mathcal{C}_{i;j}},
\end{aligned}
\end{equation}
where $R_{i\rightarrow{j}}=R_{j\rightarrow{i}}$ and $R_{i\rightarrow{l}}=R_{l\rightarrow{i}}$. Furthermore, the coding scheme includes a fixed header per transmission of $2\lceil\log_{2}2D\rceil+\lceil\log_2{h}\rceil$ bits.
\end{theorem}

\begin{figure*}[bp]
\centering
    \subfloat[Network model\label{fig:full_example_unicast_case1}]{
    \psfrag{a}[][][0.7]{$3$}
    \psfrag{b}[][][0.7]{$2$}
    \psfrag{c}[][][0.7]{$1$}
    \psfrag{d}[][][0.7]{$W^{(t)}_{2\rightarrow{1}}$}
    \psfrag{e}[][][0.7]{$W^{(t)}_{1\rightarrow{2}}$}
    \psfrag{f}[][][0.7]{$W^{(t)}_{3\rightarrow{1}}$}
    \psfrag{g}[][][0.7]{$W^{(t)}_{1\rightarrow{3}}$}
    \psfrag{i}[][][0.7]{$W^{(t)}_{3\rightarrow{2}}$}
    \psfrag{h}[][][0.7]{$W^{(t)}_{2\rightarrow{3}}$}
    \includegraphics[scale=.35]{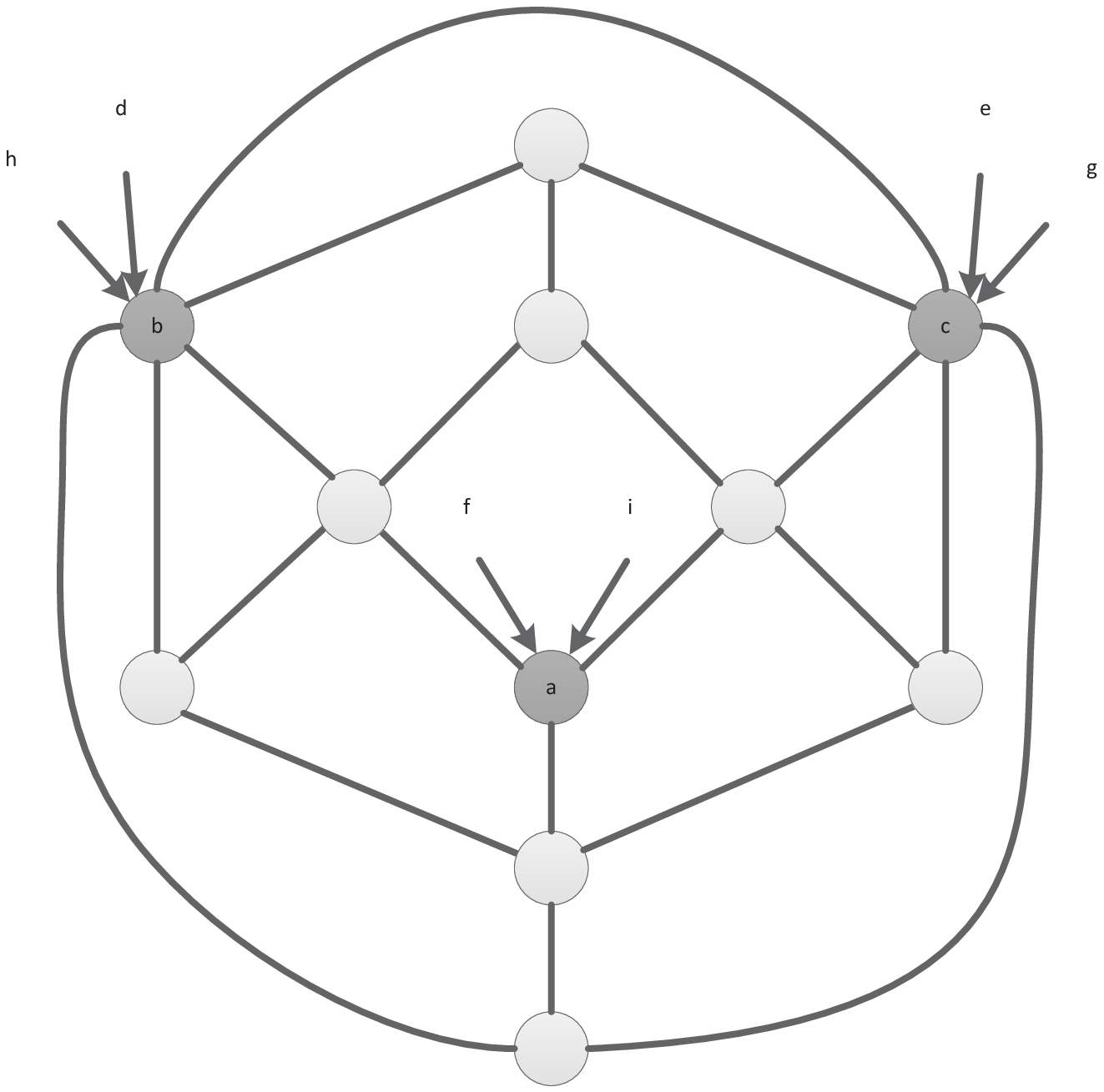}%
    }
    \hfil
    \subfloat[Flow from node $1$ to node $3$\label{fig:full_example_unicast_case2}]{
    \psfrag{a}[][][0.7]{$3$}
    \psfrag{b}[][][0.7]{$2$}
    \psfrag{c}[][][0.7]{$1$}
    \psfrag{d}[][][0.7]{$W^{(t)}_{2\rightarrow{1}}$}
    \psfrag{e}[][][0.7]{$W^{(t)}_{1\rightarrow{2}}$}
    \psfrag{f}[][][0.7]{$W^{(t)}_{3\rightarrow{1}}$}
    \psfrag{g}[][][0.7]{$W^{(t)}_{1\rightarrow{3}}$}
    \psfrag{i}[][][0.7]{$W^{(t)}_{3\rightarrow{2}}$}
    \psfrag{h}[][][0.7]{$W^{(t)}_{2\rightarrow{3}}$}
    \includegraphics[scale=.35]{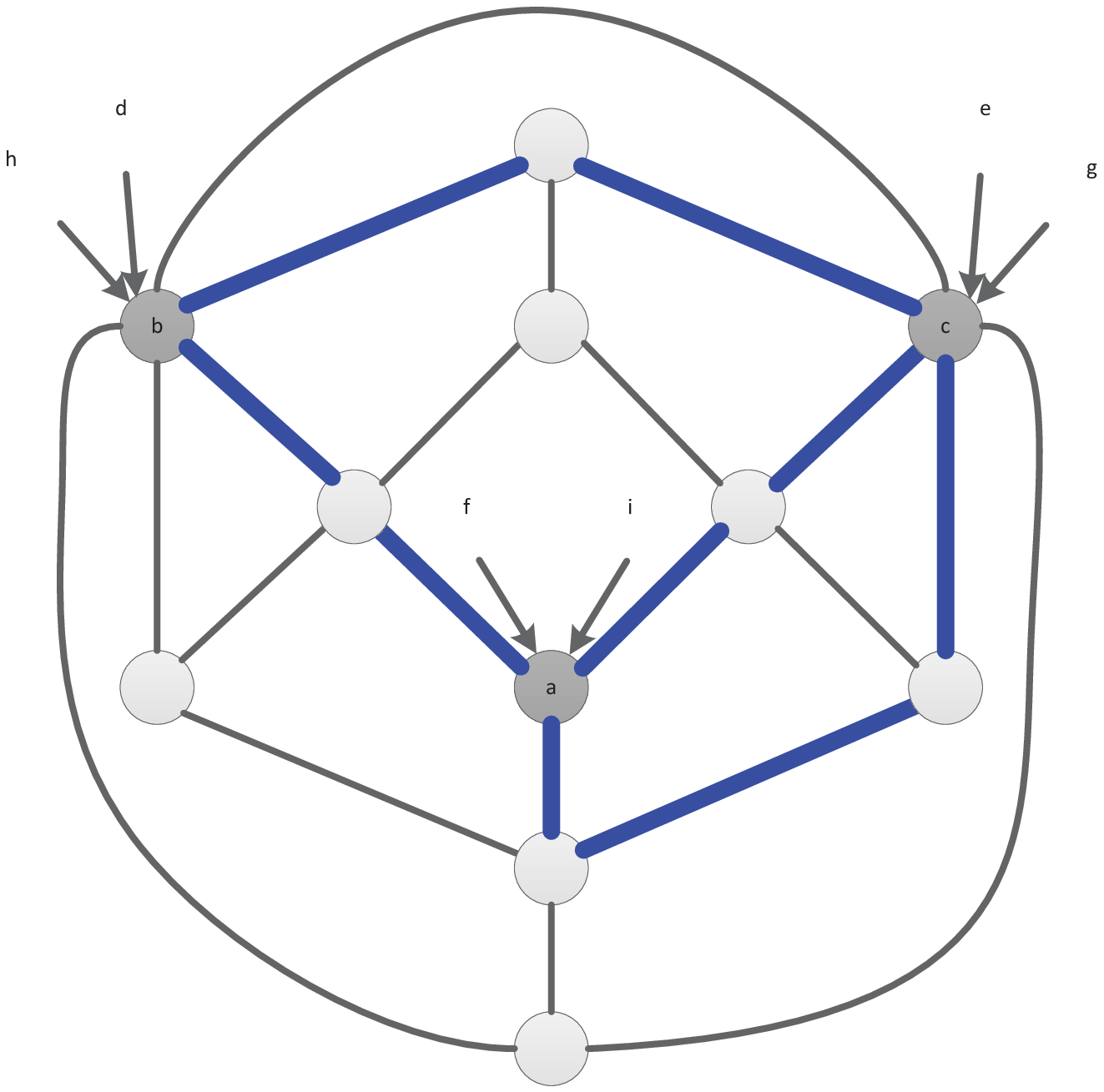}%
    }
    \hfil
    \subfloat[Flow from node $1$ to node $2$\label{fig:full_example_unicast_case3}]{
    \psfrag{a}[][][0.7]{$3$}
    \psfrag{b}[][][0.7]{$2$}
    \psfrag{c}[][][0.7]{$1$}
    \psfrag{d}[][][0.7]{$W^{(t)}_{2\rightarrow{1}}$}
    \psfrag{e}[][][0.7]{$W^{(t)}_{1\rightarrow{2}}$}
    \psfrag{f}[][][0.7]{$W^{(t)}_{3\rightarrow{1}}$}
    \psfrag{g}[][][0.7]{$W^{(t)}_{1\rightarrow{3}}$}
    \psfrag{i}[][][0.7]{$W^{(t)}_{3\rightarrow{2}}$}
    \psfrag{h}[][][0.7]{$W^{(t)}_{2\rightarrow{3}}$}
    \includegraphics[scale=.35]{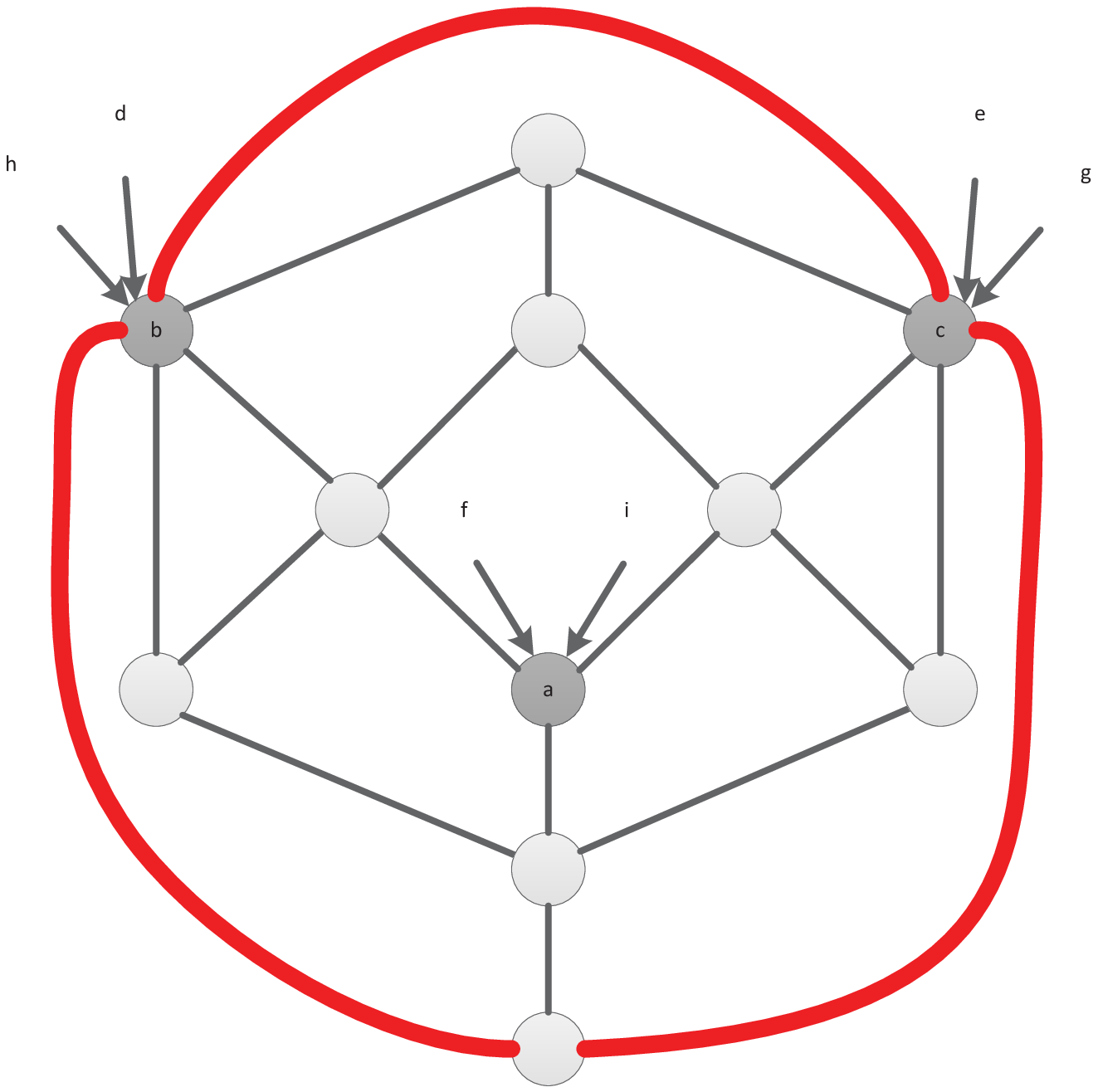}%
    }
    \caption{A network $\mathcal{G}$ is shown in \protect\subref{fig:full_example_unicast_case1}, where all the source nodes $1,2$ and $3$ communicate in a bidirectional unicast manner. In \protect\subref{fig:full_example_unicast_case2} and \protect\subref{fig:full_example_unicast_case3}, we show a corner point in the equal rate capacity region by applying the algorithm from Section \protect\ref{sec:algorithm_unicast}, where $R_{1\rightarrow{3}}=\mathcal{C}_{1;3}$, i.e., $3C$, and $R_{1\rightarrow{2}}=\mathcal{C}_{1;2,3}-\mathcal{C}_{1;3}$, i.e., $2C$.}
    \label{fig:fullExample_unicast}
\end{figure*}

\par The capacity region is upper bounded by standard minimum cut arguments. It has corner points of $(R_{1\rightarrow{2}},R_{1\rightarrow{3}},R_{2\rightarrow{3}})$ that can be found by setting the rate $R_{1\rightarrow{2}}=\mathcal{C}_{1;2}$, which yields that $R_{1\rightarrow{3}}=\mathcal{C}_{1;2,3}-\mathcal{C}_{1;2}$ and $R_{2\rightarrow{3}}=\mathcal{C}_{2;1,3}-\mathcal{C}_{1;2}$. Hence, either $R_{2\rightarrow{3}}$ or $R_{1\rightarrow{3}}$ equals zero, since node $3$ can only be in one of the cuts between nodes $1$ and $2$ and $\mathcal{C}_{1;2}$ is the minimization of all the cuts separating them. Therefore,
\begin{equation}\label{eq:min_cut_1}
\mathcal{C}_{1;2}=\min\{\mathcal{C}_{1;2,3},\mathcal{C}_{2;1,3}\}.
\end{equation}

\par A general expression for all the corner points includes $R_{i\rightarrow{j}}=\mathcal{C}_{i;j}$, $i,j\in\mathcal{S}$. Then, in case $\mathcal{C}_{i;j}=\mathcal{C}_{i;j,l}$, we get $R_{i\rightarrow{l}}=0$ and $R_{j\rightarrow{l}}=\min\{ \mathcal{C}_{j;l},\mathcal{C}_{j;i,l}-\mathcal{C}_{i;j} \}$, $l\in\mathcal{S}\setminus\{i,j\}$. Furthermore, the fact that $\mathcal{C}_{i;j,l}\leq{\mathcal{C}_{j;i,l}+\mathcal{C}_{l;i,j}}$ implies that the corner point (for the case $\mathcal{C}_{i;j}=\mathcal{C}_{i;j,l}$) can be expressed as
\begin{equation}\label{eq:gen_corner_point}
\begin{aligned}
R_{i\rightarrow{j}}&={\mathcal{C}_{i;j}}, \\
R_{j\rightarrow{l}}&={[\mathcal{C}_{j;i,l}-\mathcal{C}_{i;j}]^+}, \\
R_{i\rightarrow{l}}&=0.\\
\end{aligned}
\end{equation}

To show an achievable coding scheme for the region in (\ref{eq:unicast_upper_bound}), it is sufficient to prove that we achieve all corner points in the form of (\ref{eq:gen_corner_point}). Therefore, we present a lemma that shows how to partition each network $\mathcal{G'}$ into sub-topologies of line networks. The decomposition includes a set of disjoint paths $\mathcal{P}_{i;j}$, where $|\mathcal{P}_{i;j}|=\frac{\mathcal{C}_{i;j,l}}{C}$, and a set of disjoint paths $\mathcal{P}_{j;l}$, where $|\mathcal{P}_{j;l}|=\frac{\mathcal{C}_{j;i,l}-\mathcal{C}_{i;j,l}}{C}$ and $\mathcal{P}_{i;l}\cap{\mathcal{P}_{j;l}}=\emptyset$. By using this lemma, we can achieve the corner points in the capacity region.

\begin{lemma}\label{lemma:disjoint_unicast}
For a network $\mathcal{G'(V',E')}$ with $\mathcal{C}_{i;j}=\mathcal{C}_{i;j,l}$, there exist sets of disjoint paths $\mathcal{P}_{i;j}$ and $\mathcal{P}_{j;l}$ such that $|\mathcal{P}_{i;j}|+|\mathcal{P}_{j;l}|=\frac{\mathcal{C}_{j;i,l}}{C}$ and $|\mathcal{P}_{i;j}|=\frac{\mathcal{C}_{i;j,l}}{C}$, where $\mathcal{P}_{i;j}\cap{\mathcal{P}_{j;l}}=\emptyset$, $i,j,l\in\mathcal{S}$. Namely, $\mathcal{P}_{i;j}$ and $\mathcal{P}_{j;l}$ have no mutual edges.
\end{lemma}

\begin{IEEEproof}[Proof of Lemma \ref{lemma:disjoint_unicast}]
The maximum number of disjoint paths between nodes $i$ and $j$, $|\mathcal{P}_{i;j}|$, is $\frac{\mathcal{C}_{i;j}}{C}$ (the Max-flow Min-cut theorem \cite[Theorem 1]{cite:maximum_flow}). Since $\mathcal{C}_{i;j}=\mathcal{C}_{i;j,l}$, this is also the maximum number of paths between $i$ and $j,l$. However, in case $\mathcal{C}_{i;j,l}<\mathcal{C}_{j;i,l}$, there are more paths between $j$ and $i,l$ then $\frac{\mathcal{C}_{i;j,l}}{C}$. These paths, $\mathcal{P}_{j;l}$, are between nodes $j$ and $l$.

\par To prove that $\mathcal{P}_{i;j}$ and $\mathcal{P}_{j;l}$ are disjoint, we add $\frac{\mathcal{C}_{j;i,l}-\mathcal{C}_{i;j,l}}{C}$ direct paths between $i$ and $l$, $\mathcal{P}_{i;l}$, which yields a new network in which $\mathcal{C'}_{i;j,l}=\mathcal{C}_{j;i,l}$. As a result of (\ref{eq:min_cut_1}), we conclude that there are $\frac{\mathcal{C}_{j;i,l}-\mathcal{C}_{i;j,l}}{C}$ new paths between nodes $i$ and $j$, $\mathcal{P'}_{i;j}$. Paths $\mathcal{P'}_{i;j}$ are a union of $\mathcal{P}_{i;l}$ and $\mathcal{P}_{j;l}$, i.e., $|\mathcal{P}_{j;l}|=\frac{\mathcal{C}_{j;i,l}-\mathcal{C}_{i;j,l}}{C}$. Hence, since $\mathcal{P}_{i;j}$ and $\mathcal{P'}_{i;j}$ are disjoint according to \cite{cite:maximum_flow}, we conclude that $\mathcal{P}_{j;l}$ and $\mathcal{P}_{i;j}$ are also disjoint.
\end{IEEEproof}

\par Next, we provide the proof for Theorem \ref{theorem:unicast}.
\begin{IEEEproof}[Proof of Theorem \ref{theorem:unicast}]
Using the line topology coding scheme from Section \ref{sec:line} and Lemma \ref{lemma:disjoint_unicast}, we can achieve all the corner points in the capacity region. Specifically, we use the line topology coding scheme from Section \ref{sec:line} at each path from Lemma \ref{lemma:disjoint_unicast} to achieve a rate of $R_{i\rightarrow{j}}=\mathcal{C}_{i;j}$ and $R_{j\rightarrow{l}}=\mathcal{C}_{j;i,l}-\mathcal{C}_{i;j}$, $i,j,l\in\mathcal{S}$. Using time sharing arguments, it is straightforward to see that we achieve the capacity region under the equal rate constraint. Moreover, we obtain an asynchronous coding scheme with a maximum delay of $LD+t$ for all $t$ by using a header of $2\lceil\log_{2}2D\rceil$ bits according to the coding scheme of the line topology from Section \ref{sec:line}. Another $\lceil\log_2{h}\rceil$ bits are used to form disjoint paths in the network. Therefore, by exploiting the broadcast ability of the wireless medium, we obtain a RT NC scheme.
\end{IEEEproof}

\subsection{Algorithm to Find the Building Blocks}\label{sec:algorithm_unicast}
We introduce an optimization problem that we use to find the set of paths from Lemma \ref{lemma:disjoint_unicast}, $\mathcal{P}_{i;j}$ and $\mathcal{P}_{j;l}$, $i,j,l\in\mathcal{S}$. In this problem, we would like to find the minimum flow, $\mathbf{f}$, which maintains the condition that $|\mathcal{P}_{j;l}|=\frac{\mathcal{C}_{j;i,l}-\mathcal{C}_{i;j}}{C}$ paths from $j$ to $l$ and another $|\mathcal{P}_{i;j}|=\frac{\mathcal{C}_{i;j}}{C}$ paths from $i$ to $j$ are disjoint. Since we would like to find $\frac{\mathcal{C}_{i;j}}{C}$ paths between nodes $i$ and $j$, we demand that $\mathcal{I}_i=\frac{\mathcal{C}_{i;j}}{C}$ and $\mathcal{O}_j=\frac{\mathcal{C}_{j;i,l}}{C}$. However, for the case in which $\mathcal{C}_{i;j}<\mathcal{C}_{j;i,l}$, there are more paths emerging from node $j$. Those paths are consumed by node $l$, i.e., $\mathcal{I}_l-\mathcal{O}_l = \frac{\mathcal{C}_{j;i,l}-\mathcal{C}_{i;j}}{C}$. We guarantee that all of the paths are disjoint since $\mathbf{f}$ is a binary vector and each relay node in the network, $m\in\mathcal{V'}\setminus\mathcal{S}$, has an equal input and output flows. Hence, the outcome of this optimization problem is a binary flow that includes two sets of disjoint paths $\mathcal{P}_{i;j}$ and $\mathcal{P}_{j;l}$.

\begin{equation}
\begin{aligned}\label{eq:unicast_findLine}
\underset{{\mathbf{f}}}{\text{minimize}} & \quad\sum_{f_{(u,v)}\in\mathbf{f}}f_{(u,v)} \\
\text{subject to} & \quad\mathcal{O}_m = \sum_{n:(m,n)\in\mathcal{E'}}f_{(m,n)}, \; m = 1, \ldots, |\mathcal{V}|\\
& \quad\mathcal{I}_m = \sum_{n:(n,m)\in\mathcal{E'}}f_{(n,m)}, \; m = 1, \ldots, |\mathcal{V}|\\
& \quad\mathcal{I}_j = 0 \\
& \quad\mathcal{I}_{i} = \frac{\mathcal{C}_{i;j}}{C} \\
& \quad\mathcal{I}_l-\mathcal{O}_l = \frac{\mathcal{C}_{j;i,l}-\mathcal{C}_{i;j}}{C} \\
& \quad\mathcal{O}_j = \frac{\mathcal{C}_{j;i,l}}{C} \\
& \quad\mathcal{O}_i = 0 \\
& \quad\mathcal{O}_k = \mathcal{I}_k, \; k = 4, \ldots, |\mathcal{V}|,
\end{aligned}
\end{equation}
where $i,j,l\in\mathcal{S}$.

\subsection{Example}\label{sec:example}

\par In this subsection, we show an example of a network $\mathcal{G}$ with three source nodes that communicate in a bidirectional unicast manner (Fig. \ref{fig:fullExample_unicast}). In fact, we include an example in which we apply the algorithm from the previous Subsection \ref{sec:algorithm_unicast}, i.e., the algorithm for finding the disjoint paths between the three source nodes that, by using the line topology coding scheme from Section \ref{sec:line}, will produce a corner point in the equal rate capacity region. The corner point that is shown in \ref{fig:full_example_unicast_case2} and \subref*{fig:full_example_unicast_case3} is $(R_{1\rightarrow{2}},R_{1\rightarrow{3}},R_{2\rightarrow{3}})=(\mathcal{C}_{1;2,3}-\mathcal{C}_{1;3},\mathcal{C}_{1;3},0)$. Specifically, in this network $\mathcal{C}_{1;3}=3C$ and $\mathcal{C}_{1;2,3}=5C$, and therefore, we show three disjoint paths from node $1$ to node $3$ (Fig. \ref{fig:full_example_unicast_case2}) and another two disjoint paths from node $1$ to node $2$ (Fig. \ref{fig:full_example_unicast_case3}). The paths are disjoint since they do not pass through a common relay node or, similarly, in the equivalent graph $\mathcal{G'}$ they have no mutual edges.

\section{Simulation}\label{sec:simulation}
We simulated our results, i.e., we used a fixed number of nodes but generated a random number of edges, on an Erd{\H{o}}s-R{\'e}nyi random graph model \cite{cite:erdos}. Each edge is a bidirectional link that is drawn independently of the other links with the same probability $p$ to be with capacity $C$ or $(1-p)$ to be $0$. Then, we constructed an equivalent graph $\mathcal{G'(V',E')}$, as illustrated in Fig. \ref{fig:node_constrained}, and searched for the line and star topologies in the graphs.

\begin{figure}[h!]
\centering
    \psfrag{N}[][][0.7]{Number of Nodes}
    \psfrag{D}[][][0.7]{Size}
    \psfrag{AAA}[l][][0.7]{$\mathcal{P}_{1;3}$}
    \psfrag{BBB}[l][][0.7]{$\mathcal{P}_{1;2}$}
    \psfrag{CCC}[l][][0.7]{Sum Rate}
    \includegraphics[width=0.5\textwidth,keepaspectratio]{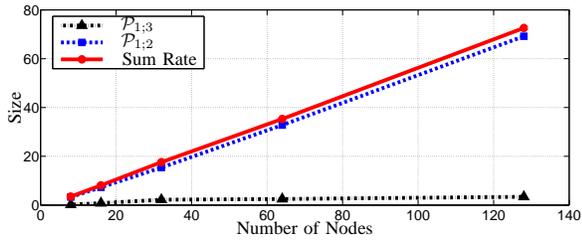}
    \caption{Simulation results of the average size of the sets $\mathcal{P}_{1;2}$ and $\mathcal{P}_{1;3}$. This simulation shows the sum rate increment that was attained using the coding scheme from Section \ref{sec:unicast}.}
    \label{fig:sim_unicast}
\end{figure}

\begin{figure}[!h]
\centering
    \psfrag{N}[][][0.7]{Number of Nodes}
    \psfrag{D}[][][0.7]{Size}
    \psfrag{AAA}[][][0.7]{$|\mathcal{R}|$}
    \psfrag{BBB}[][][0.7]{$|\mathcal{Q}|$}
    \psfrag{CCC}[][][0.7]{$h$}
    \includegraphics[width=0.5\textwidth,keepaspectratio]{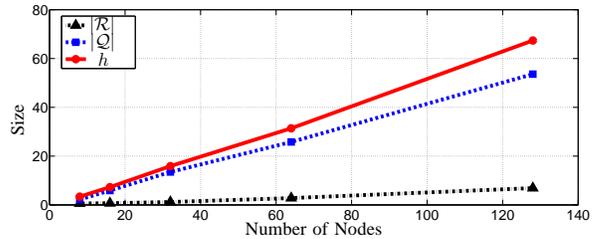}
    \caption{Simulation results of the average sizes of the sets $\mathcal{R}$ and $\mathcal{Q}$ and the upper bound $h$. Recall that $2\mathcal{R}+\mathcal{Q}\geq{h}$.}
    \label{fig:sim_multicast}
\end{figure}

\par In Fig. \ref{fig:sim_unicast}, we show a simulation result of a corner point in the equal rate capacity region of a multiple unicast network as described in Section \ref{sec:unicast}. This corner point is the maximization on the rate between nodes $1$ and $2$, i.e., $R_{1\rightarrow{2}}=\mathcal{C}_{1;2}$. Then, we search for the maximum number of disjoint paths between nodes $1$ and $3$, i.e., $R_{1\rightarrow{3}}=\mathcal{C}_{1;2,3}-\mathcal{C}_{1;2}$. Therefore, the number of paths between nodes $1$ and $2$ are likely to be greater than the number of paths between nodes $1$ and $3$. For each simulation, we compared $8$, $16$, $32$, $64$ and $128$ nodes and created ten different graphs by varying the value of $p$. We show the average number of $|\mathcal{P}_{1;2}|$ and $|\mathcal{P}_{1;3}|$ over the number of nodes in a graph $\mathcal{G}$. As expected, our result shows that each of the examined variables ($|\mathcal{P}_{1;2}|$ and $|\mathcal{P}_{1;3}|$) increases with the scale of the network, where $|\mathcal{P}_{1;2}|>|\mathcal{P}_{1;3}|$. Furthermore, this simulation shows the sum rate increment that was achieved using the coding scheme from Section \ref{sec:unicast}. By using the line topology coding scheme of Section \ref{sec:line} on each of the paths in $\mathcal{P}_{1;2}$ and $\mathcal{P}_{1;3}$, we can construct an optimal practical asynchronized coding scheme for a general network under the equal rate demand.

\par We further simulated our results on a network with the multicast demands of Section \ref{sec:multicast}, illustrated in Fig. \ref{fig:sim_multicast}. In this network, we searched for the maximum number of ring (Algorithm \ref{algo:multicast_find_line}) and line-star (Algorithm \ref{algo:multicast_find_star}) topologies. The simulation shows a comparison between $8$, $16$, $32$, $64$ and $128$ nodes. For each number of nodes, we created ten different graphs by varying the value of $p$. We show the average number of $|\mathcal{R}|$ and $|\mathcal{Q}|$ over the number of nodes in a graph $\mathcal{G}$. This simulation shows the decomposing opportunities in a graph. We recall that each ring topology contributes a rate of $C$ to each source, where each line-star topology only contributes $\frac{C}{2}$. However, we see from Fig. \ref{fig:sim_multicast} that the line-star topology makes a more significant contribution than the ring topology in the multicast problem of three users under the equal rate constraint.

\begin{figure}[!h]
\centering
    \psfrag{N}[][][0.7]{Number of Nodes}
    \psfrag{D}[][][0.7]{Transmissions}
    \psfrag{AAA}[l][][0.7]{Routing}
    \psfrag{BBB}[l][][0.7]{Coding}
    \includegraphics[width=0.5\textwidth,keepaspectratio]{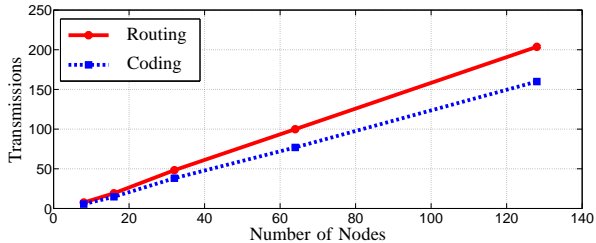}
    \caption{Simulation results of the average number of transmissions required by the relay nodes in a multicast network. As shown, we achieve better performance using the proposed coding scheme.}
    \label{fig:sim_routing_coding1}
\end{figure}

\begin{figure}[!h]
\centering
    \psfrag{N}[][][0.7]{Number of Nodes}
    \psfrag{D}[][][0.7]{Transmissions}
    \psfrag{AAA}[l][][0.7]{Routing}
    \psfrag{BBB}[l][][0.7]{Coding}
    \includegraphics[width=0.5\textwidth,keepaspectratio]{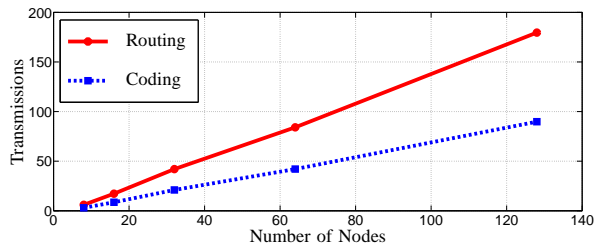}
    \caption{Simulation results of the average number of transmissions required by the relay nodes in a multiple unicast network. As shown, we achieve significantly better performance using the proposed coding scheme.}
    \label{fig:sim_routing_coding2}
\end{figure}

\par In the next simulation, Figs. \ref{fig:sim_routing_coding1} and \ref{fig:sim_routing_coding2}, we compared the performance of our coding schemes to simple routing schemes in wireless settings, $\mathcal{G(V,E)}$. Specifically, we measured the number of transmissions required by the relay nodes, $\mathcal{V}\setminus\mathcal{S}$, in the coding schemes that achieve the maximum rates for each network. The simulation shows a marked improvement in the number of required transmissions due to the efficient utilization of the wireless medium. Unlike a simple routing technique, in the proposed coding schemes we employ NC to allow at least two adjacent nodes to gain new information from each transmission. The differences between the multicast (Fig. \ref{fig:sim_routing_coding1}) and the multiple unicast (Fig. \ref{fig:sim_routing_coding2}) scenarios are due to the differences between the star topology (Section \ref{sec:star}), which makes a greater contribution in the multicast network, and the line topology (Section \ref{sec:line}), which is the only topology that is used in the multiple unicast network. Therefore, the improvement for the multiple unicast network is $100\%$, and for the multicast network is about $30\%$. Lastly, from these simulations we learned that for the same amount of information, the number of transmissions decreases significantly by using the proposed coding schemes, or equivalently, for the same number of transmissions, we achieve a higher rate.

\section{Extensions}\label{sec:extensions}
In this section, we show that the building block approach also yields a coding scheme that achieves capacity for the combined problem, namely, a network of multicast and multiple unicast demands. By using the coding schemes for the multicast (Section \ref{sec:multicast}) and the multiple unicast (Section \ref{sec:unicast}) sessions, we show that such a combination is feasible, i.e., we give a coding scheme achieving capacity. Additionally, we discuss the difficulties in the case of a network with four source nodes. Specifically, we show that the line and line-star building blocks approach that was presented in the previous sections is no longer sufficient to achieve the capacity region in the corresponding wired model.
\begin{figure}[!b]
\centering
        \psfrag{a}[][][0.8]{$1$}
        \psfrag{b}[][][0.8]{$2$}
        \psfrag{c}[][][0.8]{$3$}
        \psfrag{w}[][][0.8]{${W_{1\rightarrow{2}}^{(t)}}$}
        \psfrag{v}[][][0.8]{${W_{1\rightarrow{3}}^{(t)}}$}
        \psfrag{p}[][][0.8]{${W_{2\rightarrow{1}}^{(t)}}$}
        \psfrag{q}[][][0.8]{${W_{2\rightarrow{3}}^{(t)}}$}
        \psfrag{u}[][][0.8]{${W_{3\rightarrow{1}}^{(t)}}$}
        \psfrag{r}[][][0.8]{${W_{3\rightarrow{2}}^{(t)}}$}
        \psfrag{i}[][][0.8]{$W_1^{(t)}$}
        \psfrag{j}[][][0.8]{$W_2^{(t)}$}
        \psfrag{k}[][][0.8]{$W_3^{(t)}$}
        \includegraphics[width=2.5in]{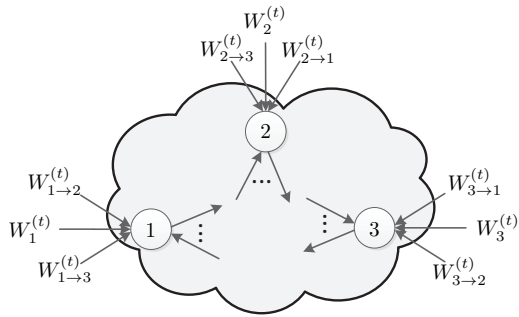}%
    \caption{Schematic illustration of the network, where $W_i^{(t)}$ is a message that node $i$ generates at time instant $t$ and that is destined to all the other source nodes, and $W_{i\rightarrow{j}}^{(t)}$ is a message that node $i$ generates at time instant $t$, and it is only destined for node $j$.}
    \label{fig:model_um}
\end{figure}

\subsection{A network with multicast and multiple unicast sessions}\label{sec:combination}
Here, we use the line and line-star topology coding schemes to show that there exists a RT coding scheme for a general wireless network, in which three source nodes communicate bidirectionally in multicast and multiple unicast manners and which achieves the equal rate capacity region (Fig. \ref{fig:model_um}). Our coding scheme achieves the capacity region under an equal rate demand. Specifically, each two-way communication is carried out at the same rate, $R_{i\rightarrow{j}}=R_{j\rightarrow{i}}$ and $R_i=R_j$, $\forall{i,j}\in\mathcal{S}$. Each source node $i\in\mathcal{S}$ produces three different messages, $W_i^{(t)},W_{i\rightarrow{j}}^{(t)}$ and $W_{i\rightarrow{l}}^{(t)}$, which are intended for the two remaining source nodes $j,l\in\mathcal{S}\setminus\{i\}$.

\begin{corollary}\label{cor:combination}
For any network $\mathcal{G'(V',E')}$ with an arbitrary delay bounded by $D$, there exists a RT coding scheme with any set of rates within the equal rate capacity region, which is
\begin{equation}\label{eq:um_upper_bound}
\begin{aligned}
2R+R_{i\rightarrow{j}}+R_{i\rightarrow{l}}&\leq{\mathcal{C}_{i;j,l}}\\
R_{i\rightarrow{j}}&\leq{\mathcal{C}_{i;j}},
\end{aligned}
\end{equation}
where $R_{i\rightarrow{j}}=R_{j\rightarrow{i}}$ and $R_{i\rightarrow{l}}=R_{l\rightarrow{i}}$, $\forall{i,j,l}\in\mathcal{S}$. Furthermore, the coding scheme includes a fixed header per transmission of $2\lceil\log_{2}2D\rceil+\lceil\log_2{h}\rceil$ bits.
\end{corollary}

\begin{IEEEproof}
The capacity region can be upper bounded by standard minimum cut arguments. For example, $\mathcal{C}_{1;2,3}$ represents the cut set bound of all the information that node $1$ receives. Nodes $2$ and $3$ transmit multicast and unicast messages to node $1$, e.g., $W^{(t)}_2,W^{(t)}_3,W^{(t)}_{2\rightarrow{1}}$ and $W^{(t)}_{3\rightarrow{1}}$. Therefore, we obtain the first upper bound in the region (\ref{eq:um_upper_bound}). The second upper bound is obtained straightforwardly by the minimum-cut maximum-flow theorem.

\par To show an achievable coding scheme for the region in (\ref{eq:um_upper_bound}), it is sufficient to prove that we achieve all corner points. There exist two different corner points: the first is the case $R_{i\rightarrow{j}}=\mathcal{C}_{i;j}$, which yields that $R=0$ since
\begin{equation}
\mathcal{C}_{i;j}=\min\{\mathcal{C}_{i;j,l},\mathcal{C}_{j;i,l}\},
\end{equation}
where $i,j,l\in\mathcal{S}$. Therefore, this corner point is strictly the multiple unicast case, which we already discussed in Section \ref{sec:unicast}. The second corner point is the case where $R=\frac{h}{2}$. In this case, there are two bidirectional unicasts sessions that equal zero. For example, assuming $h=\mathcal{C}_{1;2,3}$ yields that $R_{1\rightarrow{2}}$ and $R_{1\rightarrow{3}}$ both equal zero. The remaining unicast rate, e.g., $R_{2\rightarrow{3}}$, can be larger than zero in the general case, i.e., we can find a set of disjoint paths between nodes $2$ and $3$ such that deleting them from the network will not affect the maximum rate of the multicast session. For example, consider the network depicted in Fig. \ref{fig:example_um}. In this network, there exists a path between nodes $1$ and $2$ that does not intersect with the star topology structure, which is required to obtain the maximum multicast rate, in this case $\frac{C}{2}$.
\begin{figure}[!h]
\centering
\psfrag{a}[][][1]{$3$}
\psfrag{b}[][][1]{$1$}
\psfrag{c}[][][1]{$2$}
\psfrag{d}[][][1]{$4$}
\includegraphics[scale=0.6]{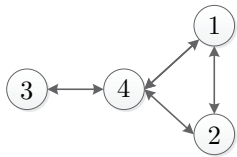}
\caption{Graph $\mathcal{G(V,E)}$, where $\mathcal{S}=\{1,2,3\}$.}
\label{fig:example_um}
\end{figure}
\end{IEEEproof}

\subsection{A network with four source nodes}\label{sec:4nodes}
\begin{figure}[!h]
\centering
    \psfrag{a}[][][1]{$1$}
    \psfrag{b}[][][1]{$2$}
    \psfrag{c}[][][1]{$3$}
    \psfrag{d}[][][1]{$4$}
    \includegraphics[scale=0.6]{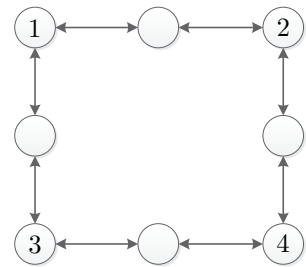}
    \caption{Multicast network $\mathcal{G(V,E)}$ with four source nodes, where a coding scheme based on the line and line-star building blocks does not achieve the equal rate upper bound.}
    \label{fig:multicast_no_building_block}
\end{figure}

In this subsection, we would like to discuss the case of a network $\mathcal{G'(V',E')}$ with four source nodes, where using the line and line-star building blocks approach failed to achieve the equal rate capacity. In the following network, nodes $1,2,3$ and $4$ communicate in a multicast manner (Fig. \ref{fig:multicast_no_building_block}). The equal rate upper bound is $R\leq\frac{2}{3}C$, which can be obtained by the standard minimum cut arguments. This upper bound can be achieved by first sending from source $i$ the message $W_i^{(t)}$ (with length $\frac{2}{3}C$) for the two adjacent nodes. Then, we use the remaining occupancy of each link to send the message $W_i^{(t)}$ to the most distinct source node from node $i$, i.e., $\frac{1}{3}C$ from each side of the node.

\par Since there is no path from node $1$ to node $4$ such that deleting it from the network will not decrease the minimum cut of node $2$, $\mathcal{C}_{2;1,3,4}$, or $3$, $\mathcal{C}_{3;1,2,4}$, there are no rings in the network. Therefore, to achieve the equal rate upper bound, two separate star topologies have to be found. However, there is only one, i.e., a rate of $\frac{1}{3}C$ is achieved. Hence, by using the building blocks of ring and line-star topologies in a network with more than three users, the equal rate upper bound is unreachable in the general case. Similarly, the ring and line-star building blocks approach also fails to achieve the equal rate capacity region for more than three users in the multiple unicast case, in which the minimum cut upper bound is not tight \cite{cite:insufficiency}. Although, our building blocks approach is insufficient in this case, we conclude that there may be another combination of building blocks that will achieve the equal rate upper bound.

\section{Conclusion}\label{sec:conclusion}
We conclude that a wireless network with three source nodes under any communication demands can be decomposed into line and line-star topologies using the building block approach. Then, by exploiting the broadcast ability of the wireless medium, we can achieve the capacity region under an equal rate constraint in the corresponding wired model. Furthermore, the coding schemes, which are based on those two canonical topologies, include many advantages, such as RT decoding delay and a small overhead. In practice, these coding schemes can be implemented on wireless networks with random transmission delays to gain better performance and power efficiency.

\bibliographystyle{IEEEtran}
\bibliography{IEEEabrv,paper_final}

\begin{thebibliography}{10}
\providecommand{\url}[1]{#1}
\csname url@samestyle\endcsname
\providecommand{\newblock}{\relax}
\providecommand{\bibinfo}[2]{#2}
\providecommand{\BIBentrySTDinterwordspacing}{\spaceskip=0pt\relax}
\providecommand{\BIBentryALTinterwordstretchfactor}{4}
\providecommand{\BIBentryALTinterwordspacing}{\spaceskip=\fontdimen2\font plus
\BIBentryALTinterwordstretchfactor\fontdimen3\font minus
  \fontdimen4\font\relax}
\providecommand{\BIBforeignlanguage}[2]{{%
\expandafter\ifx\csname l@#1\endcsname\relax
\typeout{** WARNING: IEEEtran.bst: No hyphenation pattern has been}%
\typeout{** loaded for the language `#1'. Using the pattern for}%
\typeout{** the default language instead.}%
\else
\language=\csname l@#1\endcsname
\fi
#2}}
\providecommand{\BIBdecl}{\relax}
\BIBdecl

\bibitem{cite:flow}
R.~Ahlswede, N.~Cai, S.~R. Li, and R.~W. Yeung, ``Network information flow,''
  \emph{{IEEE} Trans. Inf. Theory}, vol.~46, no.~4, pp. 1204--1216, 2000.

\bibitem{cite:field}
R.~Koetter and M.~M\'{e}dard, ``An algebraic approach to network coding,''
  \emph{{IEEE/ACM} Trans. Netw.}, vol.~11, no.~5, pp. 782--795, 2003.

\bibitem{cite:wired_model}
Y.~Wu, P.~A. Chou, Q.~Zhang, K.~Jain, Z.~Wenwu, and S.~Y. Kung, ``Network
  planning in wireless ad hoc networks: a cross-layer approach,'' \emph{{IEEE}
  J. Sel. Areas Commun.}, vol.~23, no.~1, pp. 136--150, 2011.

\bibitem{cite:code_multicast}
S.~Jaggi, P.~Sanders, P.~Chou, M.~Effros, S.~Egner, K.~Jain, and L.~M. G.~M.
  Tolhuizen, ``Polynomial time algorithms for multicast network code
  construction,'' \emph{{IEEE} Trans. Inf. Theory}, vol.~51, no.~6, pp.
  1973--1982, 2005.

\bibitem{cite:random}
T.~Ho, M.~M\'{e}dard, R.~Koetter, D.~R. Karger, M.~Effors, J.~Shi, and
  B.~Leong, ``A random linear network coding approach to multicast,''
  \emph{{IEEE} Trans. Inf. Theory}, vol.~52, no.~10, pp. 4413--4430, 2006.

\bibitem{cite:equivalent}
R.~Dougherty and K.~Zeger, ``Nonreversibility and equivalent constructions of
  multiple-unicast networks,'' \emph{{IEEE} Trans. Inf. Theory}, vol.~52,
  no.~11, pp. 5067--5077, 2006.

\bibitem{cite:linear}
S.~R. Li, R.~W. Yeung, and N.~Cai, ``Linear network coding,'' \emph{{IEEE}
  Trans. Inf. Theory}, vol.~49, no.~2, pp. 371--381, 2003.

\bibitem{cite:insufficiency}
R.~Dougherty, C.~Freiling, and K.~Zeger, ``Insufficiency of linear coding in
  network information flow,'' \emph{{IEEE} Trans. Inf. Theory}, vol.~51, no.~8,
  pp. 2745--2759, 2005.

\bibitem{cite:unicast_lo}
D.~Traskov, N.~Ratnakar, D.~Lun, R.~Koetter, and M.~Medard, ``Network coding
  for multiple unicasts: An approach based on linear optimization,'' in
  \emph{International Symposium on Information Theory (ISIT)}, 2006, pp.
  1758--1762.

\bibitem{cite:interference}
A.~Ramakrishnan, A.~Das, H.~Maleki, A.~Markopoulou, S.~Jafar, and
  S.~Vishwanath, ``Network coding for three unicast sessions: Interference
  alignment approaches,'' in \emph{Proc. Allerton Conference Communication,
  Control, and Computing}, 2010, pp. 1054--1061.

\bibitem{cite:3source}
S.~Huang and A.~Ramamoorthy, ``On the multiple-unicast capacity of 3-source,
  3-terminal directed acyclic networks,'' \emph{{IEEE/ACM} Trans. Netw.},
  vol.~22, no.~1, pp. 285--299, 2014.

\bibitem{cite:xor}
S.~Katti, H.~Rahul, W.~Hu, D.~Katabi, M.~M\'{e}dard, and J.~Crowcroft, ``Xors
  in the air: Practical wireless network coding,'' \emph{{IEEE/ACM} Trans.
  Netw.}, vol.~16, no.~3, pp. 497--510, 2008.

\bibitem{cite:coding_aware_routing}
S.~Sengupta, S.~Rayanchu, and S.~Banerjee, ``Network coding-aware routing in
  wireless networks,'' \emph{{IEEE/ACM} Trans. Netw.}, vol.~18, no.~4, pp.
  1158--1170, 2010.

\bibitem{cite:conjecture}
Z.~Li and B.~Li, ``Network coding in undirected networks,'' in \emph{Conference
  on Information Sciences and Systems (CISS)}, Princeton, New Jersey, USA,
  2004.

\bibitem{cite:unicast1}
C.~Wang, T.~Gou, and S.~Jafar, ``Multiple unicast capacity of 2-source 2-sink
  networks,'' in \emph{Global Telecommunications Conference (GLOBECOM)},
  Houston, Texas, USA, 2011.

\bibitem{cite:unicast2}
S.~Sengupta, S.~Rayanchu, and S.~Banerjee, ``An analysis of wireless network
  coding for unicast sessions: The case for coding-aware routing,'' in
  \emph{International Conference on Computer Communications (INFCOM)},
  Anchorage, Alaska, USA, 2007.

\bibitem{cite:NCvsROUTING}
K.~Jain, V.~Vazirani, and G.~Yuval, ``On the capacity of multiple unicast
  sessions in undirected graphs,'' \emph{{IEEE} Trans. Inf. Theory}, vol.~52,
  no.~6, pp. 2805--2809, 2006.

\bibitem{cite:line_dep}
M.~Bakshi, M.~Effros, W.~Gu, and R.~Koetter, ``On network coding of independent
  and dependent sources in line networks,'' in \emph{International Symposium on
  Information Theory (ISIT)}, Nice, France, 2007.

\bibitem{cite:node_constrained}
S.~M.~S. Tabatabaei~Yazdi, S.~A. Savari, and G.~Kramer, ``Network coding in
  node-constrained line and star networks,'' \emph{{IEEE} Trans. Inf. Theory},
  vol.~57, no.~7, pp. 4452--4468, 2011.

\bibitem{cite:practical}
P.~A. Chou, Y.~Wu, and K.~Jain, ``Practical network coding,'' in \emph{Proc.
  Allerton Conference Communication, Control, and Computing}, 2003.

\bibitem{cite:line}
Y.~Wu, P.~A. Chou, and S.~Y. Kung, ``Information exchange in wireless networks
  with network coding and physical-layer broadcast,'' Microsoft Research, Tech.
  Rep. MSR-TR-2004-78, 2004.

\bibitem{cite:inovite}
P.~Sadeghi, D.~Traskov, and R.~Koetter, ``Adaptive network coding for broadcast
  channels,'' in \emph{Network Coding, Theory, and Applications (NetCod)},
  Lausanne, Switzerland, 2009.

\bibitem{cite:ford}
L.~R. Ford and D.~R. Fulkerson, \emph{Flows in Networks}.\hskip 1em plus 0.5em
  minus 0.4em\relax Princeton University Press, 1962.

\bibitem{cite:karp}
R.~Karp, ``Reducibility among combinatorial problems,'' in \emph{Complexity of
  Computer Computations}, ser. The IBM Research Symposia Series, R.~Miller,
  J.~Thatcher, and J.~Bohlinger, Eds., 1972, pp. 85--103.

\bibitem{cite:approx}
N.~Garg and J.~Koenemann, ``Faster and simpler algorithms for multicommodity
  flow and other fractional packing problems,'' \emph{SIAM Journal on
  Computing}, vol.~37, no.~2, pp. 630--652, 2007.

\bibitem{cite:approx2}
N.~Garg, V.~V. Vazirani, and M.~Yannakakis, ``Primal-dual approximation
  algorithms for integral flow and multicut in trees,'' \emph{Algorithmica},
  vol.~18, no.~1, pp. 3--20, 1997.

\bibitem{cite:maximum_flow}
L.~R. Ford and D.~R. Fulkerson, ``Maximal flow through a network,''
  \emph{Canadian Math.}, vol.~8, pp. 399--404, 1956.

\bibitem{cite:erdos}
P.~Erd{\H{o}}s and A.~R{\'e}nyi, ``On the evolution of random graphs,''
  \emph{Publ. Math. Inst. Hung. Acad. Sci}, vol.~5, pp. 17--61, 1960.

\end{thebibliography}

\end{document}